\documentclass[nofootinbib,superscriptaddress,preprint]{revtex4}
\usepackage{graphicx}
\usepackage{color}
\usepackage[dvipsnames]{xcolor}

\usepackage[normalem]{ulem} 
\renewcommand\sout{\bgroup\color[rgb]{0,0,1} \ULdepth=-.5ex \ULset}

\begin{document}
%
\title{Effects of Many-body Interactions in Hypernuclei with Korea-IBS-Daegu-SKKU Functionals}
\author{Soonchul Choi}
\email{scchoi0211@ibs.re.kr}
\affiliation{Center for Exotic Nuclear Studies, Institute for Basic Science (IBS), Daejeon 34126, Korea}
\author{Emiko Hiyama}
\affiliation{Department of Physics, Tohoku University, Sendai 980-8578, Japan}
\author{Chang Ho Hyun}
\affiliation{Department of Physics Education, Daegu University, Gyeongsan 38453, Korea}
\author{Myung-Ki Cheoun}
\affiliation{Department of Physics and OMEG Institute, Soongsil University, Seoul 06978, Korea}
\date{\today}
\begin{abstract}
We investigate the properties of $\Lambda$ hyperon in $\Lambda$-hypernuclei using an effective nuclear density functional theory which is based on the low-energy effective field theory. 
It expands the energy density in the power of Fermi momentum, and consequently has multiple density dependence for the effective many-body interactions.
Starting from the effective density functional for nucleons, we determine the parameters for the 
two- and many-body $\Lambda$-$N$ interactions
added to the nucleon energy density functional by fitting to $\Lambda$-hypernuclear data.
The experimental data consist of the energy levels of a $\Lambda$ hyperon in
the $p$-, and $d$-states as well as $s$-state of $\Lambda$-hypernuclei 
in the mass range from $_\Lambda^{16}$O to $_\Lambda^{208}$Pb.
The results turn out to properly explain the data relevant to hypernuclei 
owing to the effective many-body interaction apart from a few data in light hypernuclei.
This hyperon functional is applied to study the $\Lambda$ hyperon binding energy of the neutron-rich 
$^{124-136}_\Lambda$Sn isotopes which are under consideration for the measurement at J-PARC.
Our results are shown to be insensitive to the density dependence of symmetry energy.
We also examine the nuclear matter including $\Lambda$ hyperon.
We note that the hyperon threshold density depends on the nuclear matter properties.
\end{abstract}
\maketitle

\section{Introduction}
One of ultimate goals of hypernuclear physics is to obtain information on baryon-baryon interaction 
in a unified way.
Especially, it becomes an important issue to obtain information on hyperon($Y$)-nucleon($N$) interaction. 
For this purpose, hyperon-nucleon scattering experiment is planned at J-PARC facility.
In addition, the study of structure of $\Lambda$-hypernuclei is essential for obtaining information on $YN$ interaction in nuclei.
To achieve the goal, there have been intensive efforts by high resolution
$\gamma$-ray experiments for the hypernuclei produced by many hadron probes \cite{ppnp2006,Yamamoto}. 
By combining the data and theoretical calculation such as 
shell model calculation \cite{Gal,Millener} and few-body model calculation \cite{Hiyama2009, Hiyama2012}, one can obtain constraints on the spin-independent, spin-spin, and spin-orbit terms of $\Lambda$-$N$ interaction.
Furthermore, it is necessary to study $\Lambda N$-$\Sigma N$ coupling,
which causes charge symmetry breaking effect in $\Lambda$-hypernuclei, and effective three-body force
mediated by hyperons.
Another important issue is to obtain information on short-range part of the
effective three-body force, which could contribute to the neutron matter and the core of neutron stars.
For the study of this three-body force, for instance,
M.~M. Nagels pointed out that short-range part of $\Lambda NN$ three-body force 
gave a contribution to the binding energies of $\Lambda$-hypernuclei with $A=16$ to 208 
using recent Nijmegen $YN$ potential, Extended Soft Core 16 potential model \cite{Nagels}.
At J-PARC facility, it is planned to perform systematic
measurement of binding energies with wide mass region by energy spectroscopy \cite{Aoki:2021cqa}.
Considering this situation, it is required to perform energy spectra
of $\Lambda$-hypernuclei over wide mass region using reliable models
and to study the short-range part of $\Lambda NN$ three-body force.

Along this line, Skyrme force model provides a simple but efficient and robust platform for the access to the problem.
In the Skyrme force model, strong forces are approximated to the low energy limit,
in which two- and three-nucleon forces could be represented in terms of $\delta$ functions.
Transforming the three-body force to the form of two-body forces, one obtains a term that depends explicitly on density.
In the standard Skyrme force, there was only one density-dependent term that is originated from the effective three-body force.
But it is questionable whether a single density-dependent term would be 
sufficient for a proper description of the many-body effects.
In addition, the original form of the density-dependent term is proportional to density, but in later works the power of the density
has taken 
numbers smaller than one (e.g. 1/3 or 1/6) to satisfy certain constraints or obtain better results from the fitting.
For example, in the development of the SLy model, compression modulus of the symmetric matter $K_0$ is assumed
$210 \pm 20$ MeV, and correlation between $K_0$ and the power of the density-dependent term ($\sigma$)
was explored \cite{Chabanat:1997qh}.
It is shown that small $\sigma$ gives small $K_0$, and $\sigma=1/6$ reproduces best the assumed range of $K_0$.
In spite of these ambiguity and arbitrariness, Skyrme force models have been successful in describing the properties of
numerous nuclei, from light to heavy, and also useful for the nuclei
from the valley of stability to the drip lines.
Standard Skyrme force, i.e. a potential with only one density-dependent term has also been applied to $\Lambda$-hypernuclei
\cite{guleria2012, hiyama2014}, and the corresponding theoretically obtained results are in good agreement to the data of single-particle levels of the 
$\Lambda$ hyperon over a wide range of mass number.
However, the issue of arbitrariness in the density-dependent term has not been argued and explored seriously yet.

In the present work, we revisit the interaction of $\Lambda$ hyperons in nuclear medium at finite density and zero temperature
within a non-relativistic framework of nuclear density functional theory.
(There are also numerous works based on the relativistic mean field theory.
For recent works, see Refs.~\cite{rmf1, rmf2, rmf3}). 
In the KIDS (Korea-IBS-Daegu-SKKU) density functional formalism, similar to the low-energy effective field theory for few-nucleon systems, the energy per particle stemming from the strong interaction is expanded in the power of $k_F/m_\rho$ where 
$k_F$ and $m_\rho$ are the Fermi momentum and rho-meson mass, respectively. 
At zero temperature $k_F \propto \rho^{1/3}$ where $\rho$ is the matter density,
so the expansion in terms of $k_F$ gives rise to multiple density dependence in the in-medium nuclear potential.
The expansion rule of the KIDS density functional eliminates the arbitrariness of the density-dependent term
in the standard Skyrme force.
In addition, the expansion scheme allows one to account for many-body effects beyond the mean field approximation,
determine the optimal number of necessary terms to describe finite nuclei and infinite nuclear matter correctly,
estimate the uncertainty of theoretical results and predictions, and identify the range of density at which the application of the model is valid.
From an extensive analysis, it was shown that the hierarchical priority is established in accordance with the order counting,
and the optimal number of terms for both nuclei and nuclear matter is seven \cite{kids-nm, kids-nuclei2}.
It's been also shown that with the seven terms nuclear matter equation of state (EoS) could be described consistently with
existing data and other theoretical calculations over a density range from as low as $\sim 0.01 \rho_0$ that corresponds to the dilute neutron matter 
\cite{kids-nuclei1} to as high as $\sim 6 \rho_0$ which is thought to be a typical density at the center of most massive neutron star \cite{kids-ksym}, 
where $\rho_0$ denotes the nuclear saturation density.

Employing the same expansion rule to the in-medium interaction of $\Lambda$ hyperons,
at first we consider the number of density-dependent terms necessary to describe the experimental data sufficiently accurately.
As a measure for the agreement to data, we consider the mean deviation (MD) from experimental data.
If the objective function for fitting, MD, gets saturated after using certain number of terms in the density expansion, the number may correspond to the optimal number of necessary terms.
After the optimal number is determined, we consider the uncertainty in the density dependence of the nuclear symmetry energy in determining the $\Lambda$ hyperon interactions in nuclear medium. 
As a result, several sets of parameters for the interaction of $\Lambda$ hyperons are obtained by fitting the $\Lambda$ hyperon interactions to the $\Lambda$-hypernuclear data in the mass range $A=16-208$.
Models thus obtained are applied to the prediction of the properties of
light hypernuclei such as $^8_\Lambda$He, $^9_\Lambda$Li, $^{10}_\Lambda$B, $^{11}_\Lambda$ B, $^{12}_\Lambda$B,
$^{12}_\Lambda$C, and $^{13}_\Lambda$C.
Application is further extended to the single-$\Lambda$ Sn isotopes
because the effect of $\Lambda NN$ three-body force and the dependence on the symmetry energy are likely to appear more clearly in the heavier neutron-rich $\Lambda$-hypernuclei,
and Sn has a lot of neutron-rich isotopes. 
Therefore, if we could see contribution of the $\Lambda NN$ three-body effect  
in the single-$\Lambda$ hypernuclear isotopes of Sn nuclei,
it would be meaningful for experimentalists to observe them. 

We outline the work as follows.
In Sect. II, we present the strategy how models and the parameters therein are constrained, and determine the number 
of terms that optimizes the description of $\Lambda$ hyperon interactions in nuclei. 
In Sect. III, we determine models within the uncertainty of the symmetry energy,
and apply them to investigate the properties of light hypernuclei and single-$\Lambda$ hypernuclei in the isotopic chain of Sn nucleus.
In Sect. IV, we calculate the particle fraction of the $\beta$-equilibrium and charge-neutral infinite nuclear matter,
and investigate the effect of saturation properties on the creation of $\Lambda$-hyperon in the neutron star core.
In Sect. V, we summarize the work.

\section{Formalism}

In the KIDS formalism, energy density functional is expanded in powers of Fermi momentum, or equivalently $\rho^{1/3}$
where $\rho$ is the baryon density.
Functionals for the nucleon are described and explained in detail in the previous publications 
(see e.g. Ref. \cite{kids-nuclei2}).
The total energy density of the KIDS model for hypernuclei is written as
\begin{equation}
    H_{\text{total}} = H_{N} + H_{\Lambda}.
\end{equation}
The $H_{N}$ is given as follows:
\begin{eqnarray}
 H_{N} &=& \frac{\hbar^2}{2 m_N} \tau_N  + \frac{3}{8} t_0 \rho^2_N
-\frac{1}{8}{ t_0(1+2x_0)}\rho_N^2\delta^2
\nonumber \\ & & 
+\frac{1}{16}\sum_{n=1}^{M_s}t_{3n}\rho^{2+n/3}
-\frac{1}{48}\sum_{n=1}^{M_a}{ t_{3n}(1+2x_{3n})}\rho^{2+n/3}_N\delta^2 
\nonumber \\ & &
+\frac{1}{64}(9t_1-5t_2-4t_2 x_2)(\nabla\rho_N)^2
- \frac{1}{64}(3t_1+6{ t_1 x_1}+t_2+ 2 t_2 x_2){(\nabla\rho_N\delta)^2}
\nonumber \\ & & 
+\frac{1}{8}(2t_1+ t_1 x_1 + 2t_2+  t_2 x_2)\tau \rho_N
- \frac{1}{8}(t_1+  t_1 x_1 - t_2-  2t_2 x_2)\sum_q{\rho_q\tau_q}
\nonumber \\ & & 
+ \frac{1}{2} W_0\left(\vec{\nabla}\rho_N\cdot\vec{J}_N+\sum_q\vec{\nabla}\rho_q\cdot\vec{J}_q\right)
\label{eq:N}
\end{eqnarray} 
where $\delta$, $\tau$, and $\vec{J}$ are isospin asymmetry, $\delta=(\rho_n-\rho_p)/\rho_N$, kinetic energy density and current density, respectively.
In the interaction of nucleons, density-dependent terms are summed up to $M_s=3$ and $M_a=4$.
For the hyperon part, it can be separated into two terms, which are two-body and many-body terms,
\begin{equation}
    H_{\Lambda}=H_{\Lambda N} + H_{\Lambda \rho}.
    \label{eq:lambda}
\end{equation}
For the many-body interactions of $\Lambda$ hyperons such as $\Lambda NN$, we assume a similar expansion in terms of the nucleon density as
\begin{eqnarray}
H_{\Lambda \rho} &=& \frac{3}{8} \rho_\Lambda \sum_{i=1}^{N_f} u_{3i} \left(1+\frac{1}{2} y_{3i} \right) \rho_N^{1+i/3}. 
\label{eq:lambda-rho}
\end{eqnarray}
This $\Lambda \rho$ term is an extention and generalization of the
density-dependent three-body interaction in the Skyrme force \cite{Chabanat:1997qh}.
This new type of functional is added to the two-body interaction terms given as
\begin{eqnarray}
H_{\Lambda N} &=& \frac{\hbar^2}{2 m_\Lambda} \tau_\Lambda + u_0 \left(1+\frac{1}{2} y_0 \right) \rho_N \rho_\Lambda \nonumber \\
& & + \frac{1}{4} (u_1+u_2) (\tau_\Lambda \rho_N + \tau_N \rho_\Lambda) 
+ \frac{1}{8} (3u_1 -u_2) (\vec{\nabla} \rho_N \cdot \vec{\nabla} \rho_\Lambda)
\nonumber \\
& & + \frac{1}{2} W_\Lambda (\vec{\nabla} \rho_N \cdot \vec{J}_\Lambda 
+ \vec{\nabla} \rho_\Lambda \cdot \vec{J}_N).
\label{eq:lambda-N}
\end{eqnarray}
Model parameters $u_0$, $y_0$, $u_1$, $u_2$, $W_\Lambda$, $u_{3i}$ and $y_{3i}$ are fitted to $\Lambda$-hypernuclear data.
In this study, {$\Lambda$-binding energies are taken from $\Lambda$-hypernuclei data, which are given by \cite{guleria2012}
\begin{equation}
    B_\Lambda = E_N^{\text{core}}-E_\Lambda=\int d^3 r H_{\text{total}}(Z,N,\Lambda=0) - \int d^3 r H_{\text{total}}(Z,N,\Lambda=1).
\end{equation}
    
The available data are single-$\Lambda$ hypernuclear data over the mass number from $_\Lambda^{8}$He to $_\Lambda^{208}$Pb.
Since the mean field approximation is appropriate for the large mass number systems, 
we consider the data of masses above $^{16}_\Lambda$O.
In Tab.~\ref{tab:exp}, we summarize the $\Lambda$-hypernuclear data used in fitting the model parameters.
Data of light nuclei are available, but in order to examine the predictive power of the theory, 
they and the energy levels at $f$ and $g$ orbitals in heavy nuclei
will be used in the comparison with the prediction of the present approach.

\begin{table}
\begin{center}
\begin{tabular}{llllll}\hline
Nuclei \phantom{aa}& $1s$ (MeV) & $1p$ (MeV) & $1d$ (MeV) & \\ \hline
$^{16}_{\Lambda}{\rm O}$ \cite{prl2004, npa1998} 
& $12.50 \pm 0.35$ \phantom{aa} & & & & \\
$^{28}_{\Lambda}{\rm Si}$ \cite{ppnp2006, prc1996} 
& $16.60 \pm 0.20$ & $7.0 \pm 0.2$  & & & \\
$^{32}_{\Lambda}{\rm S}$ \cite{plb1979}
& $17.50 \pm 0.50$ & & & & \\
$^{40}_{\Lambda}{\rm Ca}$ \cite{ptp1994, npa1988}
& $18.70 \pm 1.1$ & & & & \\
$^{51}_{\Lambda}{\rm V}$ \cite{prl1991, prc2001}
& $19.9 \pm 1.0$ & & & & \\
$^{89}_{\Lambda}{\rm Y}$ \cite{ppnp2006, prc2001}
& $23.10 \pm 0.50$ & $16.50 \pm 4.1$ & $9.1\pm 1.3$ & & \\
$^{139}_{\Lambda}{\rm La}$ \cite{ppnp2006}
& $24.50 \pm 1.20$ & $20.40 \pm 0.6$ & $14.3\pm0.6$ & &  \\
$^{208}_{\Lambda}{\rm Pb}$ \cite{ppnp2006}
& $26.30 \pm 0.80$ & $21.90\pm 0.6$ \phantom{aa} & $16.8\pm0.7$ & & \\
\hline
\end{tabular}
\end{center}
\caption{
Experimental data of the binding energy of a $\Lambda$ hyperon in nuclei used in fitting the in-medium 
interaction of $\Lambda$ hyperon given by Eqs. (\ref{eq:lambda-rho}) and (\ref{eq:lambda-N}).
In the notation of nuclei, mass number includes the contribution of $\Lambda$ hyperon.
For example, $^{16}_\Lambda$O means $Z=8$, $N=7$, and $\Lambda=1$.}
\label{tab:exp}
\end{table}

In the KIDS formalism, higher order contributions are added systematically, 
so it is feasible to determinine the number of terms that reproduce the nuclear data optimally.
In order to find the optimal number of terms for the interaction of the $\Lambda$ hyperon in nuclear medium,
we consider the number of terms $N_f$ in Eq. (\ref{eq:lambda-rho}) from one to five.
Each model labeled by Y$N_f$ is fitted to the data in Tab.~\ref{tab:exp}. 
It is important to define the proper objective function to optimize the model parameter \cite{Dobaczewski:2014jga}.
Most essential concern of the work is search for methodology to improve the description of hyperon interaction in many-body system in a systematic way.
To achieve this goal, the accuracy is measured by the mean deviation value from experimental data, hereafter referred to as MD, defined by
\begin{equation}
    \textmd{MD} = \frac{1}{N_d}\sum_{i=1}^{N_d}\left( \frac{M_i^{\rm exp}-M_i^{\rm cal}}{\sigma_i}\right)^2,
\end{equation}
where $N_d$ and $\sigma_i$ are number of data and experimental errors in Tab.~\ref{tab:exp}, respectively.

\begin{figure}[ht!]
\begin{center}
\includegraphics[width=9.5cm]{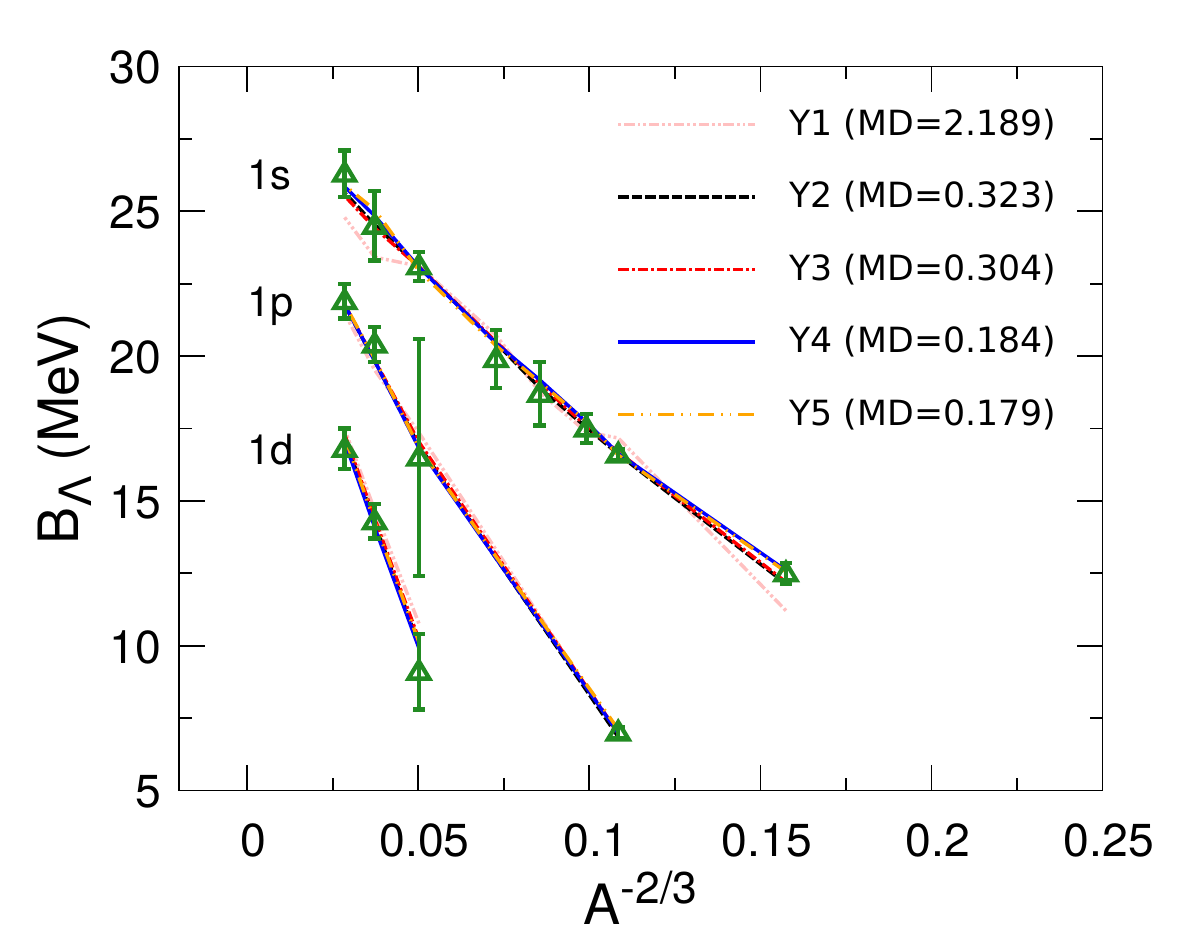}
\end{center}
\caption{Result of fitting to the $\Lambda$-binding energy data in Tab.~\ref{tab:exp} with the Y$N_f$ ($N_f=1-5$) model for the $\Lambda\rho$ interaction and the KIDS0 model for the nucleon.}
\label{fig:1}
\end{figure}

Figure~\ref{fig:1} shows the result of fitting together with the data adopted in the fitting \cite{Agrawal:2006ea}.
In this fitting procedure, we use KIDS0 for the nucleon part, $H_N$ \cite{kids-nuclei1}.
Numbers in the parenthesis denote the MD values.
As it has to be, accuracy of the model is improved as the number of terms is increased.
However rather than a monotonic behavior, MD decreases stepwise.
Improvement from Y1 to Y2 is dramatic, and Y2 and Y3 reproduce the data at a similar accuracy.
Agreement to data becomes more accurate in Y4, and there is a small difference between Y4 and Y5.
The models having multiple $\Lambda \rho$ interactions (Y2 -- Y5) reproudce the data
within the experimental uncertainty ranges. 
In the case of nucleon, we learned that the optimal number of terms is three for the symmetric nuclear matter and four for the asymmetric nuclear matter \cite{kids-nm, kids-nuclei2}.
We find that the $\Lambda$-$N$ interaction has similar numbers for the optimized description.
In the application of the model afterwards, we will consider only the four-term model (Y4) for the interaction of $\Lambda$ hyperon.

Nuclear symmetry energy is an essential issue for understanding the nuclear properties
in the vicinity of neutron-rich region and the EoS of dense nuclear matter.
Density dependence of the symmetry energy is conventionally expanded around the saturation density $\rho_0$ as
\begin{eqnarray}
S(\rho) = J + L x + \frac{1}{2} K_{\rm sym} x^2 + O(x^3),
\end{eqnarray}
where $x=(\rho - \rho_0)/3 \rho_0$.
In the numerical calculation, we assume the saturation density, $\rho_0 = 0.16\,{\rm fm}^{-3}$.
Behavior around $\rho_0$ is most sensitive to the value of $L$, slope parameter, and then $K_{\rm sym}$, the curvature of symmetry energy.
These parameters are critical in determining the structure of neutron-rich nuclei,
and bulk properties of the neutron star, but uncertainty is still large.
In a recent work, several sets of symmetry energy parameters $J$, $L$, $K_{\rm sym}$ have been derived \cite{kids-k0}, which satisfy both nuclear data and neutron star observation which are mass $2.14^{+0.10}_{-0.09} \;M/M_\odot$ at 68.3\% credibility interval (MSP J0740+6620 ) \cite{NANOGrav:2019jur} and radius $12.71^{+1.14}_{-1.19}$ at 68\% credibility interval (PSR J0030+0451) \cite{Riley:2021pdl}.
Five models labeled KIDS0, and KIDS-A, B, C, D in Tab.~\ref{tab:matter-properties} have different values of $J$, $L$, and $K_{\rm sym}$ as well as the compression modulus $K_0$ \cite{kids-k0}.
Using these models, one can explore the effect of uncertainties in the EoS of nuclear matter to the in-medium interaction of $\Lambda$ hyperons.
It has been shown that the creation of $\Lambda$ hyperons in the core of a neutron star, and consequential bulk properties of neutron stars depend strongly on the EoS of nucleonic matter \cite{ijmpe2015}.
With the number of density-dependent $\Lambda\rho$ terms fixed to four, for each KIDS model, we will determine the parameters of the $\Lambda$-$N$ interaction by fitting them to the data in Tab.~\ref{tab:exp}.

\begin{table}
\begin{center}
\begin{tabular}{lccccc}\hline\hline
\phantom{aa}& KIDS0 \phantom{a} & KIDS-A \phantom{a} & KIDS-B \phantom{a} & KIDS-C \phantom{a} & KIDS-D \phantom{a} \\ \hline
$K_0$ & 240 & 230 & 240 & 250 & 260 \\
$J$ & 32.8 & 33 & 32 & 31 & 30 \\
$L$ & 49.1  & 66 & 58 & 58 & 47 \\
$K_{\rm sym}$ & $-156.7$ & $-139.5$ & $-162.1$ & $-91.5$ & $-134.5$ \\
\hline\hline
\end{tabular}
\end{center}
 \caption{Nuclear matter properties at the saturation density for KIDS models \cite{kids-k0}.
All the models have identical saturation density 0.16 fm$^{-3}$ and the binding energy per nucleon 16 MeV.
Compression modulus $K_0$, and the symmetry energy parameters $J$, $L$, and $K_{\rm sym}$
are in the unit of MeV.}
\label{tab:matter-properties}
\end{table}

\section{ Many-body Effects on the Hyperon in Nuclei}

\begin{table}
    \begin{center}
    \begin{tabular}{cccccc}\hline\hline
    \phantom{aa}& KIDS0 \phantom{aa} & KIDS-A \phantom{aa} & KIDS-B \phantom{aa} & KIDS-C \phantom{aa} & KIDS-D \phantom{aa} \\ \hline
    $u_0$\phantom{aa}  & $-160.95264$ & $-128.81502$ & $-169.22118$ & $-125.07505$ & $-125.94104$ \\
    $u_1$\phantom{aa}   & 70.76000  & 94.95210 & 56.69966 & 85.71671 & 75.82594 \\
    $u_2$\phantom{aa}  & $-15.02857$  & $-5.14575$ & $-12.34702$ & $-7.23919$ & $-11.74793$ \\
    $u_{31}$\phantom{aa} & $-422.21514$ & $-398.48003$ & $-371.23408$ & $-231.60546$ & $-206.39651$ \\
    $u_{32}$\phantom{aa} & 373.02533  &  265.47960 & 222.62548 & 276.26060 & 359.13476 \\
    $u_{33}$\phantom{aa} & 248.20334  & 276.00074  & 220.71935 & 208.07528 & 190.90642 \\
    $u_{34}$\phantom{aa} & $-304.90644$ & $-298.23538$ & $-114.46241$ & $-299.70071$ & $-270.48269$ \\
    $y_0$\phantom{aa}    & 3.95994    & 5.19600 & 3.67720 & 5.75962 & 5.33203 \\
    $y_{31}$\phantom{aa} & $-6.60578$  & $-7.65872$ & $-8.29438$ & $-9.53907$ & $-8.17689$ \\
    $y_{32}$\phantom{aa} & 5.65621    & $-1.45742$ & 9.06089 & 7.47400 & 9.44507 \\
    $y_{33}$\phantom{aa} & $-0.89714$   & 6.40191 & $-2.14865$ & 7.10293 & $-4.59458$ \\
    $y_{34}$\phantom{aa} & $-2.15451$   & $-5.52651$ & $-4.17324$ & 2.09727 & $-0.53146$ \\ \hline
    MD \phantom{aa} & 0.1844 & 0.2096 & 0.2229 & 0.2085 & 0.2093 \\
    \hline\hline
    \end{tabular}
    \end{center}
    \caption{Parameters of the Y4 model obtained from fitting to the data in Tab.~\ref{tab:exp}. Units are 
MeV fm$^3$ for $u_0$, MeV fm$^5$ for $u_1$ and $u_2$, and MeV fm$^{3+i}$ for $u_{3i}$. $y_0$ and $y_{3i}$ are dimensionless.}
\label{tab:para}
\end{table}

For a consistent determination of the $\Lambda$-$N$ interaction,
parameters in the Y4 interaction are fitted to the data in Tab.~\ref{tab:exp} for each KIDS model.
Results for the parameters and the MD are summarized in Tab.~\ref{tab:para}.
As a whole the parameters are similar to each other, but a few notable differences also exist.
Magnitude of $u_{31}$ differs by a factor of 2. 
Negative sign implies attractive force.
If a coefficient in the potential energy is negative and large in magnitude,
it makes the chemical potential of $\Lambda$ hyperon small,
and it leads to early onset of the hyperon creation, which subsequently makes the equation of state soft and causes the hyperon puzzle in neutron star \cite{Choi:2020dbh}.

In order to have better understanding of the behavior of $\Lambda$-$N$ potential,
we consider the Hamiltonian density given by Eq.~(\ref{eq:lambda-rho})
and the term proportional to $u_0$ in Eq.~(\ref{eq:lambda-N}) divided by $\rho_\Lambda$;
\begin{eqnarray}
h_3 (\rho_N) &=& \frac{H_{\Lambda\rho}}{\rho_\Lambda} 
= \frac{3}{8} \sum_{i=1}^N u_{3i} \left(1+\frac{1}{2} y_{3i} \right) \rho_N^{1+i/3}, \nonumber \\
h_0 (\rho_N) &=& u_0 \left(1+ \frac{1}{2} y_0 \right) \rho_N.
\label{eq:h03}
\end{eqnarray}
Since $u_1$ and $u_2$ are smaller than $u_0$ and $u_{3i}$ by orders of magnitude, we omit those terms in the consideration.
\begin{figure}[t]
\begin{center}
\includegraphics[width=5.3cm]{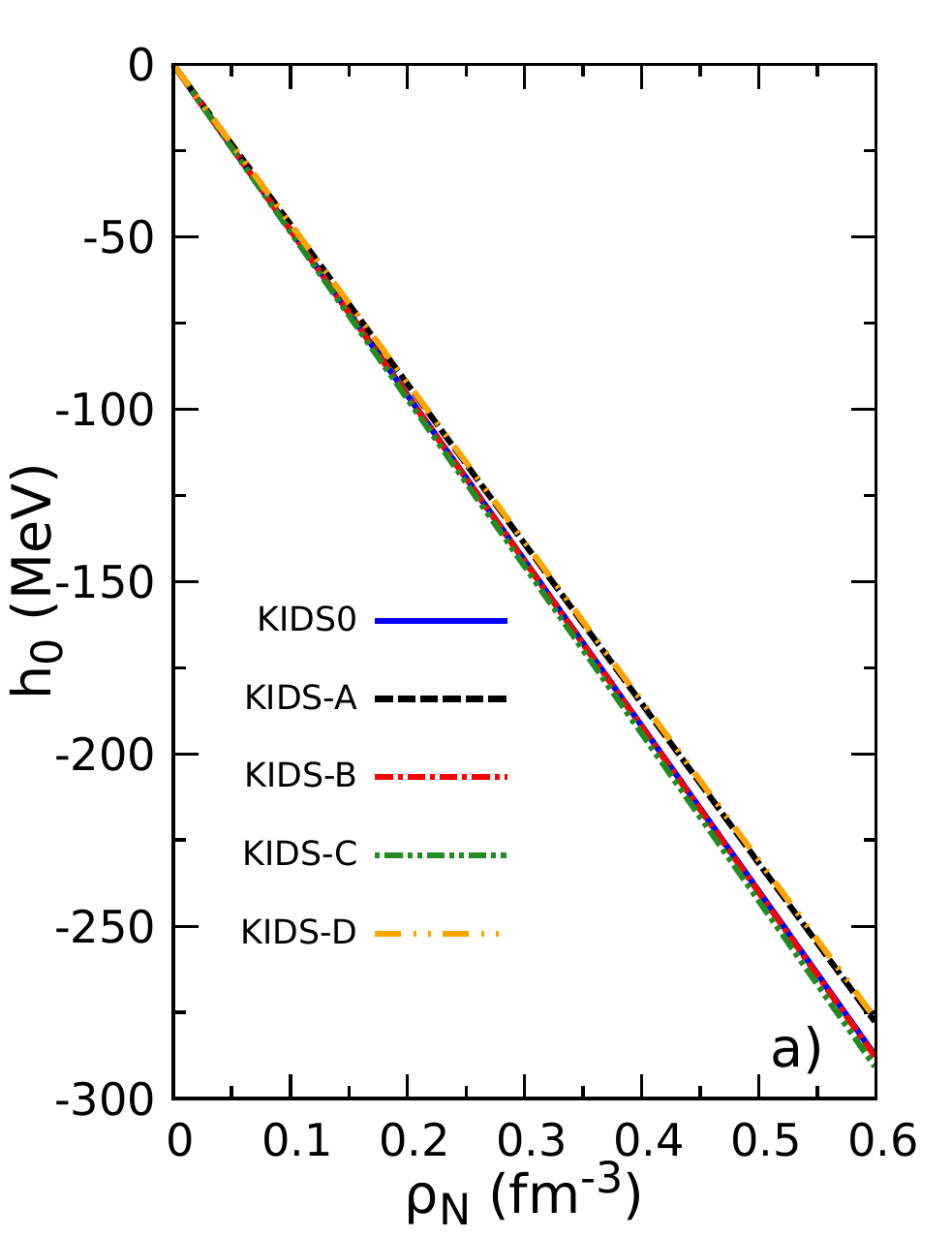}
\includegraphics[width=5.3cm]{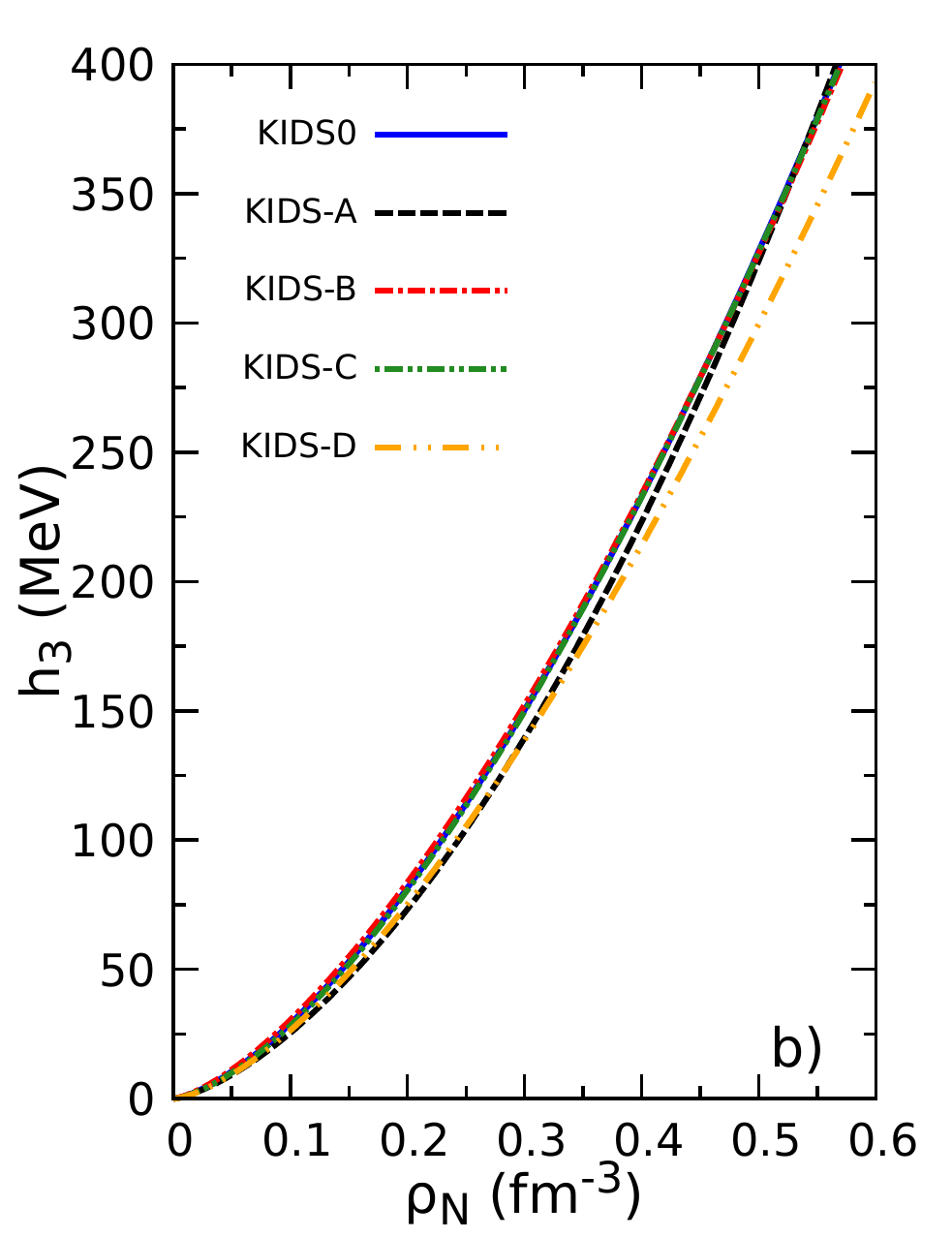}
\includegraphics[width=5.3cm]{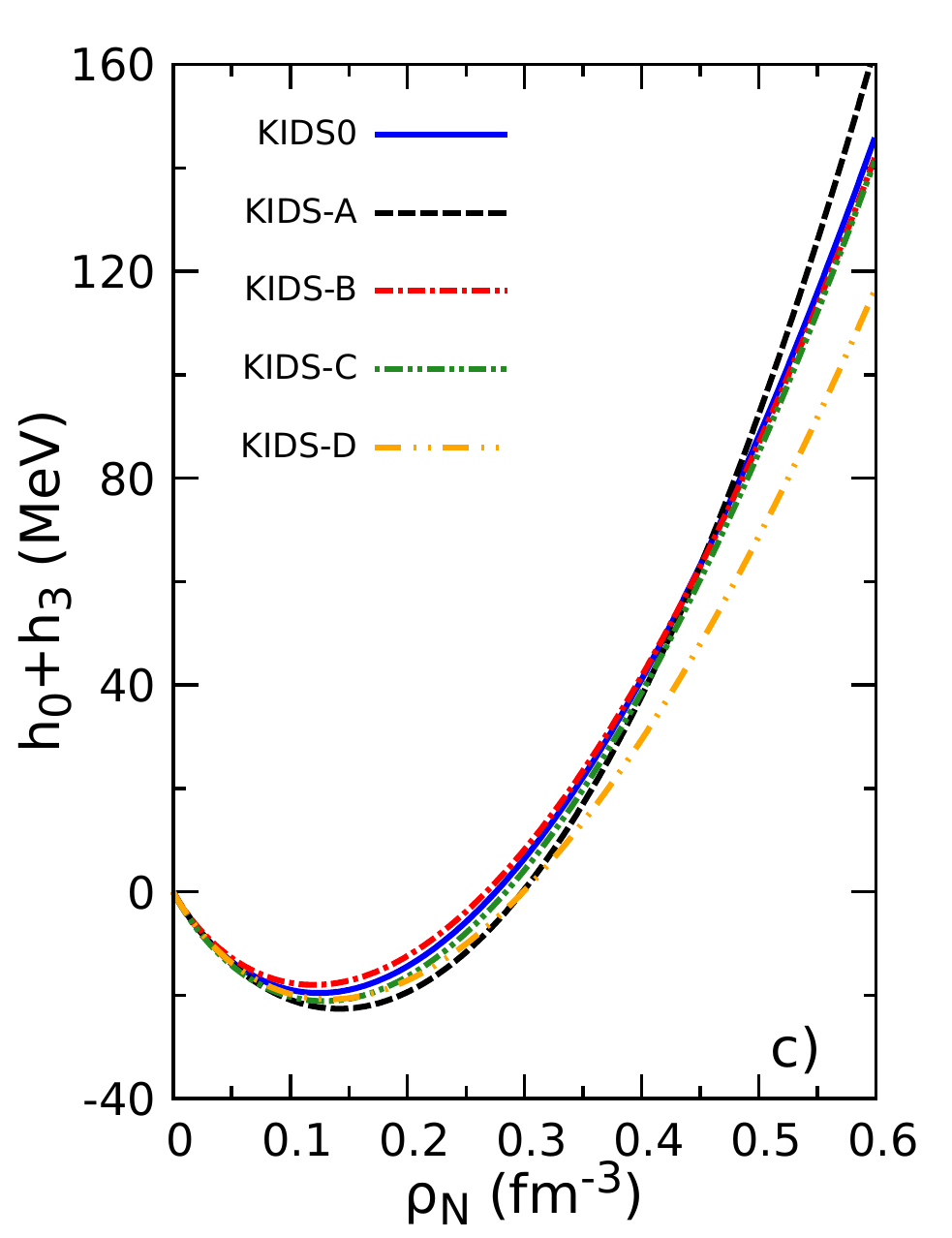}
\end{center}
\caption{Density dependence of the $\Lambda$-$N$ potential in Eq.~(\ref{eq:h03}). (a): $h_0$, (b): $h_3$ defined in Eq.~(\ref{eq:h03}), and (c): $h_0 + h_3$ 
as a function of the nucleon density $\rho_N$.}
\label{fig2}
\end{figure}
The results corresponding to the Eq.~(\ref{eq:h03}) are displayed in Fig.~\ref{fig2}.
The models show similar behavior for the two-body contribution $h_0$ around the saturation density, 
but sum of $h_0$ and $h_3$ discloses non-negligible model dependence.
$\Lambda$-$N$ potential depth at $\rho_0$ is about $-30$ MeV, so it is consistent with the depth of empirical optical potential.
Behavior of $h_0+h_3$ in $\rho_N = 0.3-0.5\, {\rm fm}^{-3}$ is relevant to the hyperon puzzle in the neutron star matter
because many models predict that hyperons begin to show up in the core of neutron stars in this density region.
Hyperons in the neutron star matter will be considered in the next section.

At densities above $0.5\, {\rm fm}^{-3}$, models again show manifest difference.
KIDS-A model increases fastest, KIDS0, B, C models are similar, and KIDS-D model is by far softer than the other models.
Softness of the KIDS-D model could be understood from the sign of $u_{33}$ term: sign of the term in the KIDS-D model is negative, but it is positive in the other models.
Stiffness at this high density could have significant effect on the particle fraction in the neutron star core,
which affects the stiffness of EoS, and consequently the mass-radius relation and the maximum mass.
We leave detailed discussions for neutron stars as a future work.

\begin{figure}
\begin{center}
\includegraphics[width=9.5cm]{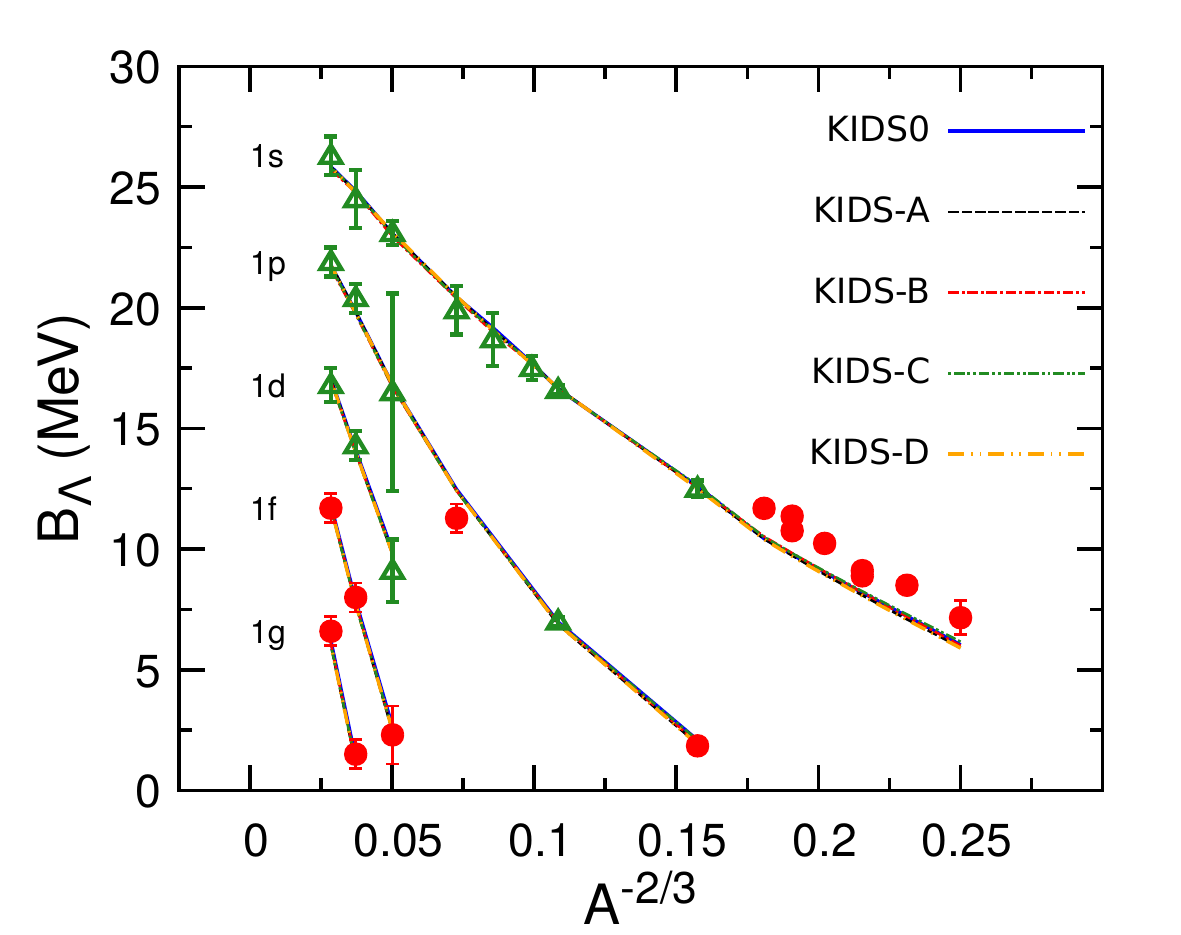}
\end{center}
\caption{Same as Fig.~\ref{fig:1}, but for the KIDS0, KIDS-A, KIDS-B, KIDS-C and KIDS-D models with Y4 hyperon interaction
together with the data used in the fitting (green triangles) and those not used (red circles).}
\label{fig3}
\end{figure}

Next we examine the performance of models by applying them to hypernuclear properties that are not included in the fitting data.
Results are summarized in Fig.~\ref{fig3}.
Lines denote the theoretical results and empty green triangles represent the data used in the fitting.
Data unemployed in the fitting process are the levels of nuclei lighter than 
$^{16}_\Lambda$O \cite{npa2005, prl1995, acta2004, prc2003, prl2001, prc2002, prl2009}, 
and the levels in the $f$- and $g$-orbitals of heavy nuclei in Refs. \cite{ppnp2006, prc2001}.
They are marked in filled red circles.
Theory lines overlap so closely that it is not necessary to distinguish models when nuclear properties are referred to.
This coincidence could be understood from the similarity of MD values in Tab.~\ref{tab:para}.

Present models reproduce the input data well within the error bars.
In the $1f$ and $1g$ states, model predictions are also within the experimental uncertainties.
However, comparison with the levels in the light nuclei is not as good as heavy nuclei.
Models consistently underestimate the experimental data.
Fitting with light nuclei as well as heavy ones was performed in Refs.~\cite{guleria2012, hiyama2014}.
When the light nuclei data are accounted in the fitting, results for the light nuclei become better than this work.
In that case, however, theory does not reproduce a few data for heavy nuclei.
For example, HP$\Lambda2$ model in Ref.~\cite{guleria2012} gave a result out of the data of $1s$ state in 
$^{208}_\Lambda$Pb ($A^{-2/3} = 0.0285$).
There was a mismatch with data of $^{28}_\Lambda$Si ($A^{-2/3} = 0.108$) in the $1s$ state \cite{hiyama2014}.
The problem in light nuclei could be explored in detail in a separate work.

In the comparison with the existing single-$\Lambda$ hypernuclear data, the dependence on the symmetry energy turns out to be very weak as shown in Fig.~\ref{fig3}.
But, symmetry energy could have more evident effect to the properties of neutron-rich nuclei such as binding energy, size and shape \cite{nd2022}.
Therefore, the measurement of $\Lambda$ hyperon binding energy in the neutron-rich isotopes could show more sensitivity to the symmetry energy.
In addition, exploration of the order-by-order contribution will shed light on the significance of high order terms and convergence behavior of the expansion scheme for the $\Lambda$ hyperon interaction.
Indeed, effect of the symmetry energy in the neutron-rich region of Sn isotopes has been explored in the KIDS model by considering neutron drip line \cite{kids-k0} and the neutron skin thickness \cite{kids-nskin}.
Both works showed that the properties of Sn isotopes in the neutron-rich region are sensitive to the density dependence of the symmetry energy.
Therefore, we consider the single-$\Lambda$ Sn isotopes $^A_\Lambda$Sn in the mass range $A=124-136$.

\begin{figure}
\begin{center}
\includegraphics[width=0.49\textwidth]{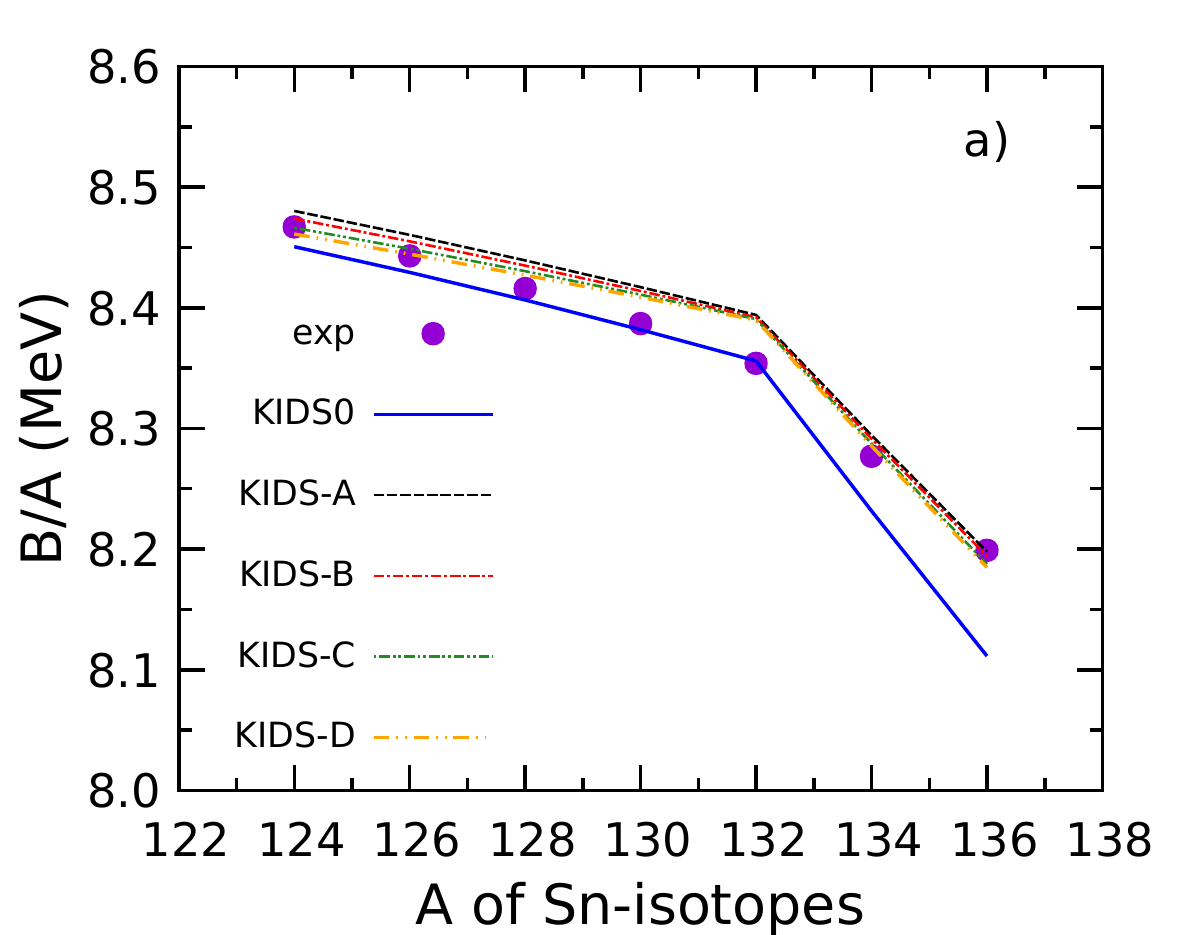}
\includegraphics[width=0.49\textwidth]{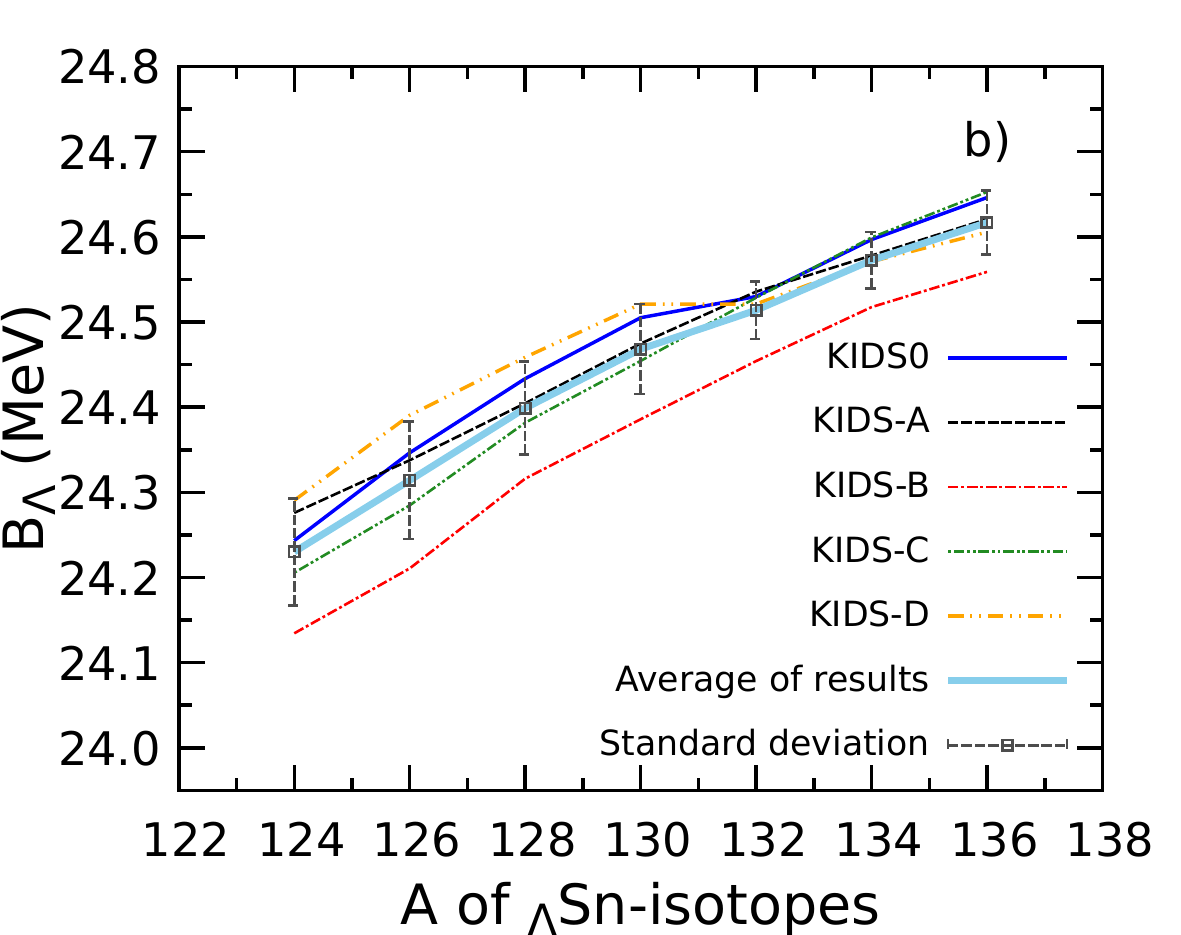}
\end{center}
\caption{(a): Binding energy per nucleon of Sn isotopes. Model results are drawn in lines, and experimental values are 
denoted with circle. (b): Prediction of the $\Lambda$ hyperon binding energies in the $_\Lambda$Sn isotopes. Mean values and possible error bars due to theoretical models are also presented.}
\label{fig4}
\end{figure}

Fig.~\ref{fig4}-(a) shows the binding energy per nucleon of non-hyperonic Sn isotopes
$B/A=(NM_n+ZM_p-M(N,Z))/A$ obtained from the KIDS models.
Experimental data are taken from Ref.~\cite{Audi:2002rp}.
KIDS-A,B,C,D models show very similar behavior, and KIDS0 is distinguished from them.
KIDS0 model agrees to experimental data better than the other models for $A=124-132$,
but shows larger discrepancy from data for $A=134-136$.
Largest difference between the result of KIDS0 model and experimental data happens at $A=136$, 
which is about 1\% of the experimental value.
For the KIDS-A,B,C,D models, maximum difference is 0.6\% for $A=132$.
As a whole, the five KIDS models reproduce the data at sufficiently high accuracy.
Fig.~\ref{fig4}-(b) shows the prediction for the binding energy of a $\Lambda$ hyperon bound in Sn isotopes.
Binding energy increases monotonically, and the shape could be approximated with a linear function.
The biggest difference within the models occurs at $A=126$ between KIDS-B and KIDS-D.
The difference is 0.18 MeV, so it is about 0.7\% of the binding energy of the KIDS-B model.
For the theoretical predictions, we also show the mean and error bars by the considered models in Fig.~\ref{fig4}-(b).

\begin{figure}
    \begin{center}
    \includegraphics[width=0.49\textwidth]{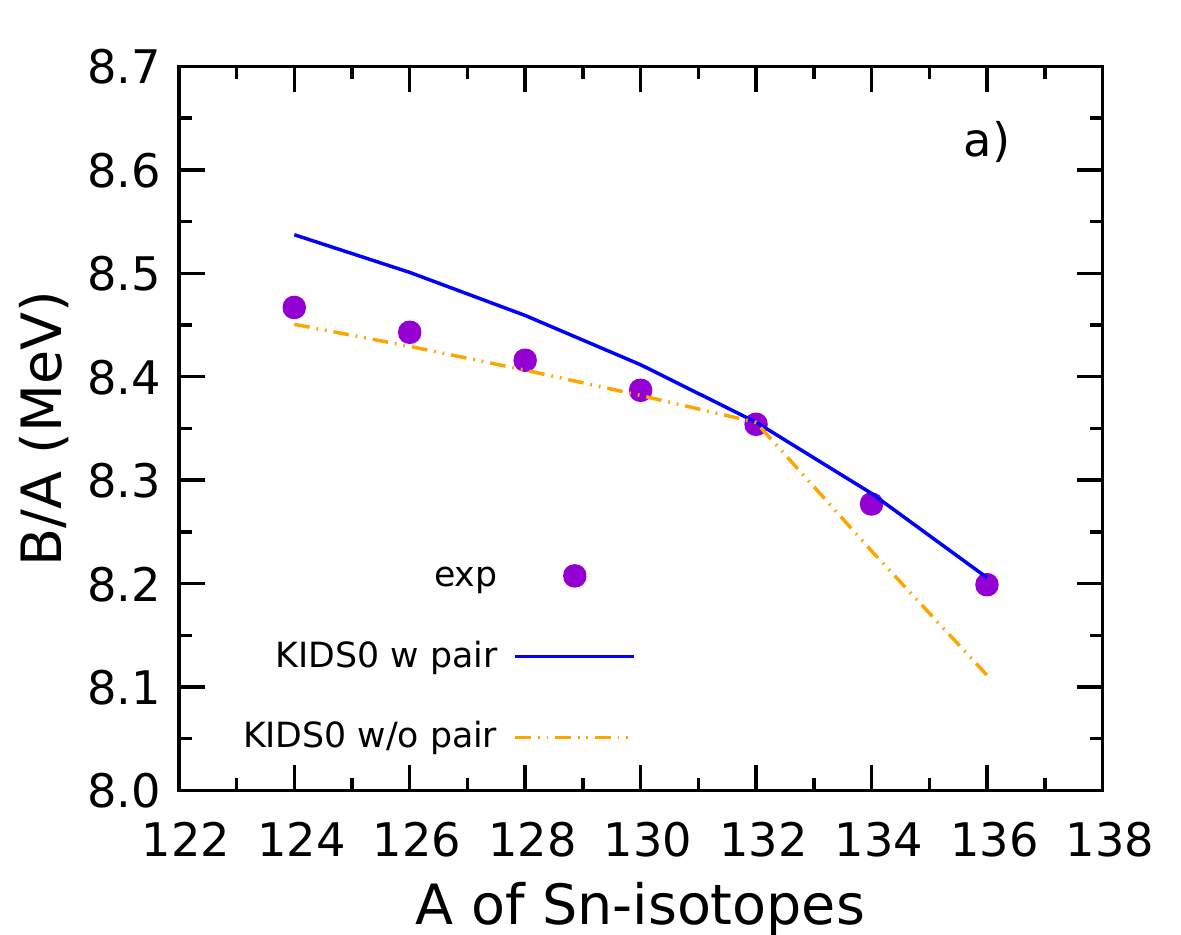}
    \includegraphics[width=0.49\textwidth]{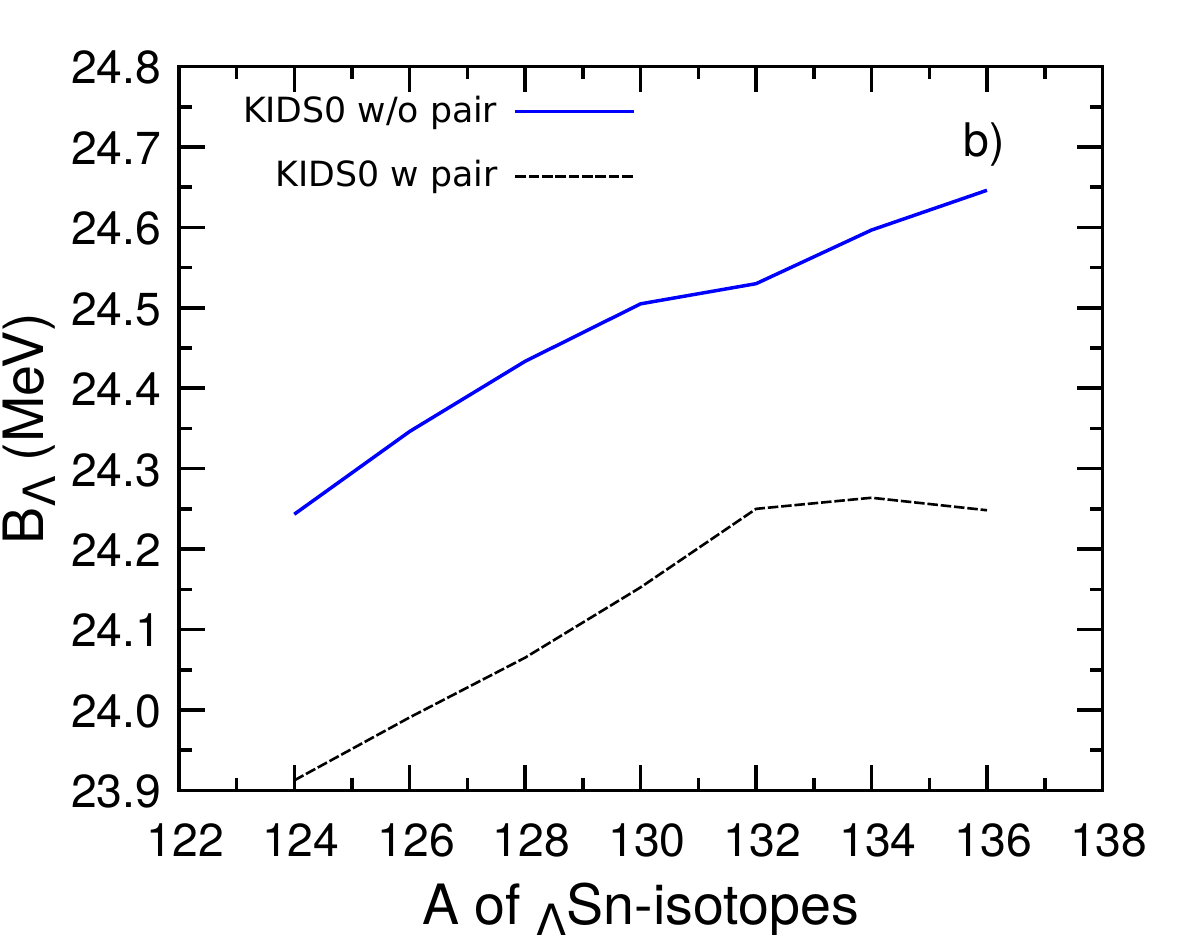}
    \end{center}
    \caption{Same as Fig.~\ref{fig4} but the case with and without pairing effect for the KIDS0 model.}
    \label{fig5}
\end{figure}

The nucleonic pairing effect is considered in Fig.~\ref{fig5}. 
To access the role of pairing interactions, we perform the parameter fitting with pairing forces.
A model for the pairing interaction is employed as follows:
\begin{equation}
    E_{\text{pair}}=\sum_{q\in p,n}G_q\left[ \sum_{\alpha\in q}\sqrt{\nu_\alpha(1-\nu_\alpha)}\right]^2,
\end{equation}
where $G_q$ stands for pairing strength which are taken to be constant for $q\in n,p$ ($G_q=29/A$ MeV for each case) and the occupation probability of state $\alpha$ is denoted by $\nu_\alpha$ \cite{Chandel:2003jt}.
In Fig.~\ref{fig5}-(a), we compare the binding energy of non-hyperonic Sn isotopes with and without the pairing forces.
For $A<132$, the result with pairing overestimates the data, but the agreement to data becomes accurate for $A \geq 132$.
Maximum difference from data is about 0.7 MeV for the result with pairing.
Similary, Fig.~\ref{fig5}-(b) discloses that the $\Lambda$-hyperon binding energies are affected within 0.35 MeV. From the result, it is verified that the binding energy of $\Lambda$ in Sn-isotope hypernuclei is not sensitive to the pairing interaction because its contribution is within 1.5\% of binding energy.

\begin{figure}
\begin{center}
\includegraphics[width=9.5cm]{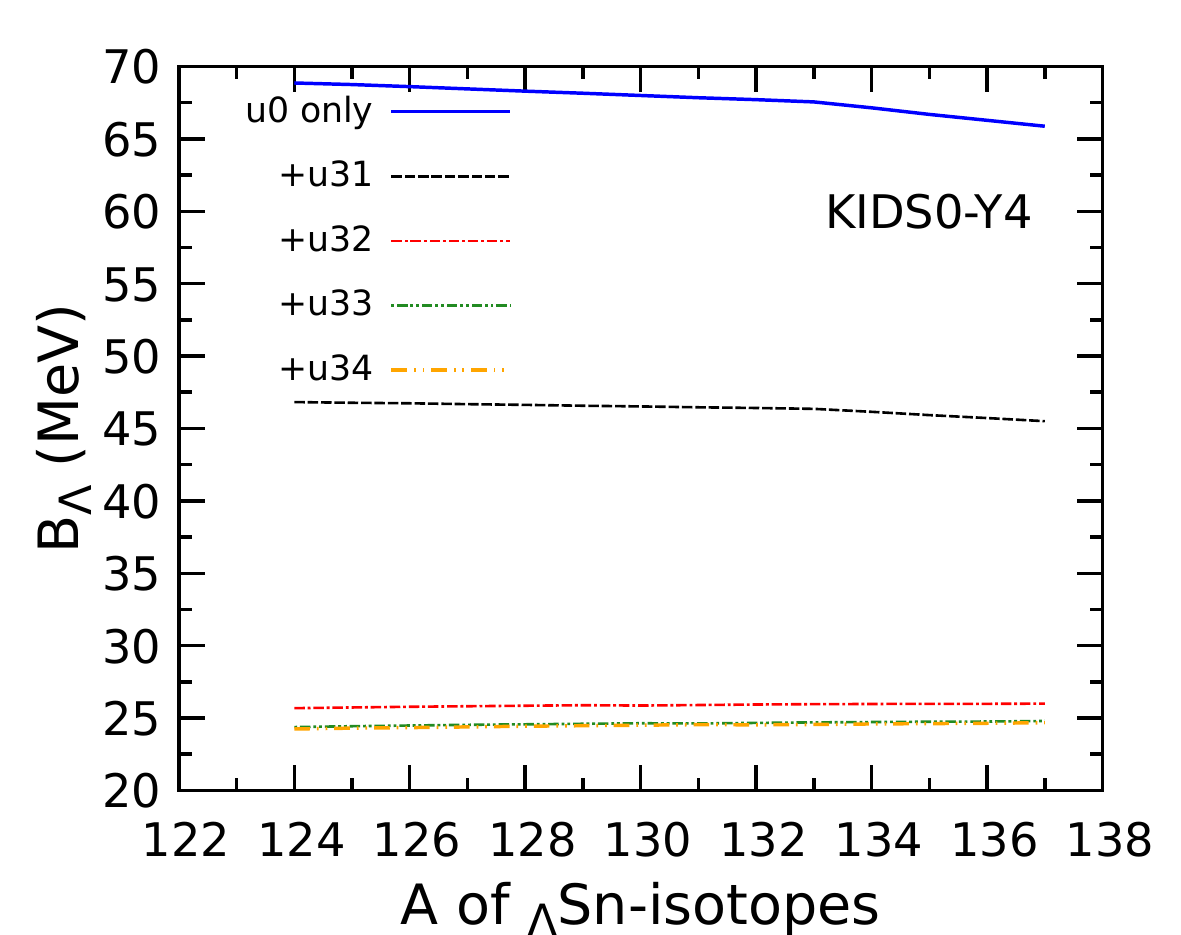}
\end{center}
\caption{Role of the density-dependent terms in the binding energy of $\Lambda$ hyperon in the $_\Lambda$Sn isotopes.}
\label{fig6}
\end{figure}

The idea to expand the energy density in the power of momentum was inspired by the expansion rules in the low-energy effective field theory.
The expansion is assumed to be perturbative,
and the perturbative behavior of the theory could be examined by comparing the contributions from each order.
Figure~\ref{fig6} shows the binding energy calculated with only the $u_0$ term at first, 
and higher order terms are added in the increasing order for the KIDS0 model.
With only $u_0$ term, $\Lambda$ hyperon is strongly over bound in nuclei.
When a repulsion by the $u_{31}$ term is added, attraction by the $u_0$ term is significantly reduced,
but it is not sufficient to obtain the physical value.
Realistic binding energy is achived when the $u_{32}$ term is added to the prior $u_0 + u_{31}$ potential.
In terms of the magnitude, each $u_{31}$ and $u_{32}$ term reduces the binding energy by about 23 MeV, respectively.
The change originated from the next higher order $u_{33}$ term is only about 2 MeV.
The result of full potential (line denoted by +u34) is hardly distinguishable from the precedent order (line denoted by +u33),
which demonstrates that the contribution of higher order terms is highly suppressed.
The result confirms that the multiple density dependence of the EDF is a necessary condition
for a correct and converging description of $\Lambda$-hypernuclei in the neutron rich domain.

\section{$\Lambda$-hyperon in the $\beta$-equilibrium matter}

$\Lambda$ hyperons embedded in hypernuclei are surrounded by nucleons at densities close to the nuclear saturation.
Therefore, hypernuclear data are valuable in suggesting constraints relevant to the $\Lambda$-$N$ interactions at the saturation density.
The models calibrated to the hypernuclei data are expected to describe well the hypernuclear phenomenology at the saturation density,
and in addition applicable to densities below or above the saturation density to a certain extent.
Neutron star provides an ideal laboratory for probing the validity of hypernuclear models at densities well above the saturation density.
The application to neutron star is especially important for exploring a solution to the hyperon puzzle problem in the neutron star \cite{Choi:2020dbh}.

Equation of state for the matter in the core of neutron stars is obtained as a function of the particle density.
Density of each particle is obtained as solutions of the charge neutrality
\begin{equation}
\rho_p = \rho_e + \rho_\mu,
\end{equation}
and chemical equilibrium
\begin{eqnarray}
    \mu_n = \mu_p + \mu_{e^-},
    \nonumber \\
    \mu_\Lambda = \mu_n,
    \nonumber \\
    \mu_{\mu} = \mu_{e^-},
    \label{eq:chemeq}
    \end{eqnarray}
where $\mu_i$ denotes the chemical potential of particle $i$.

\begin{figure}
\begin{center}
\includegraphics[width=7cm]{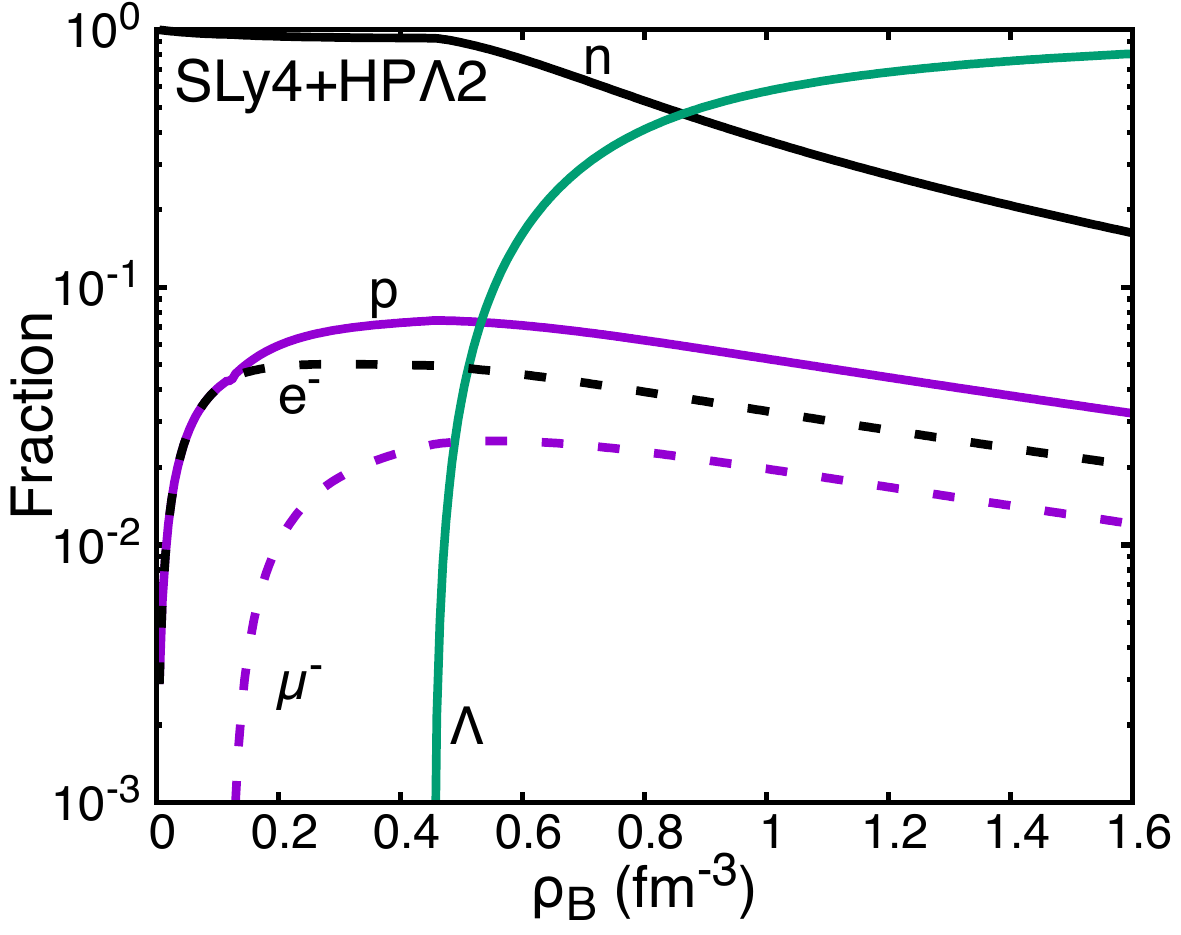}
\includegraphics[width=7cm]{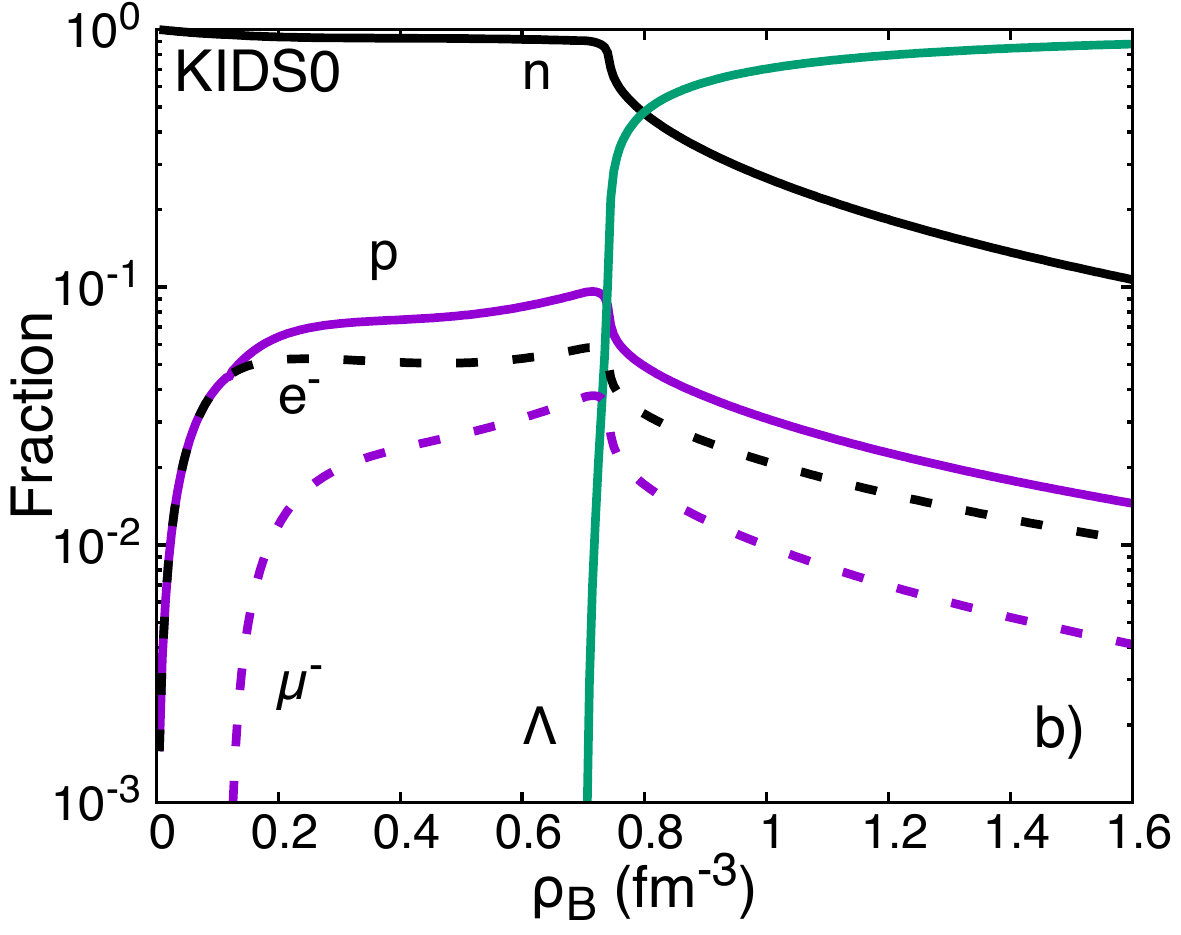}
\includegraphics[width=7cm]{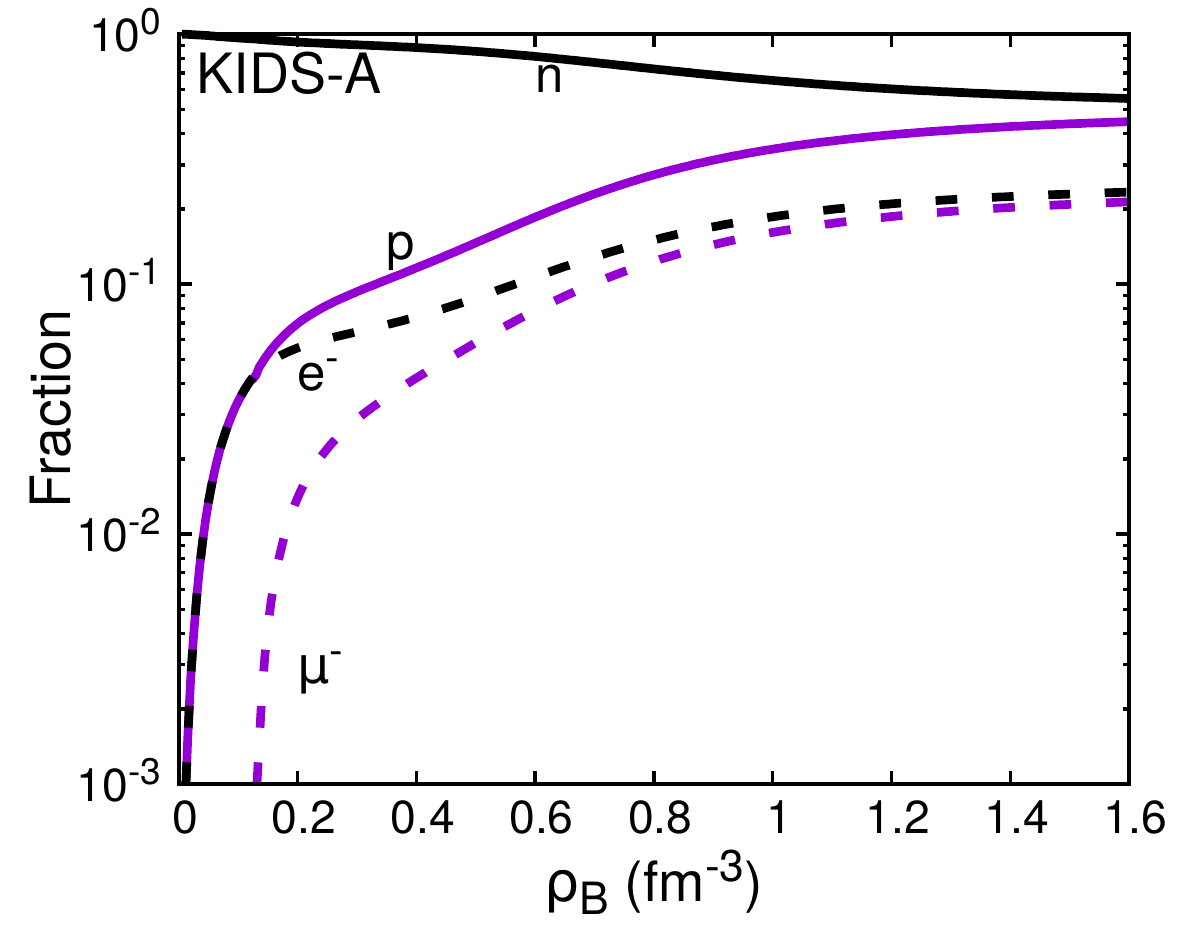}
\includegraphics[width=7cm]{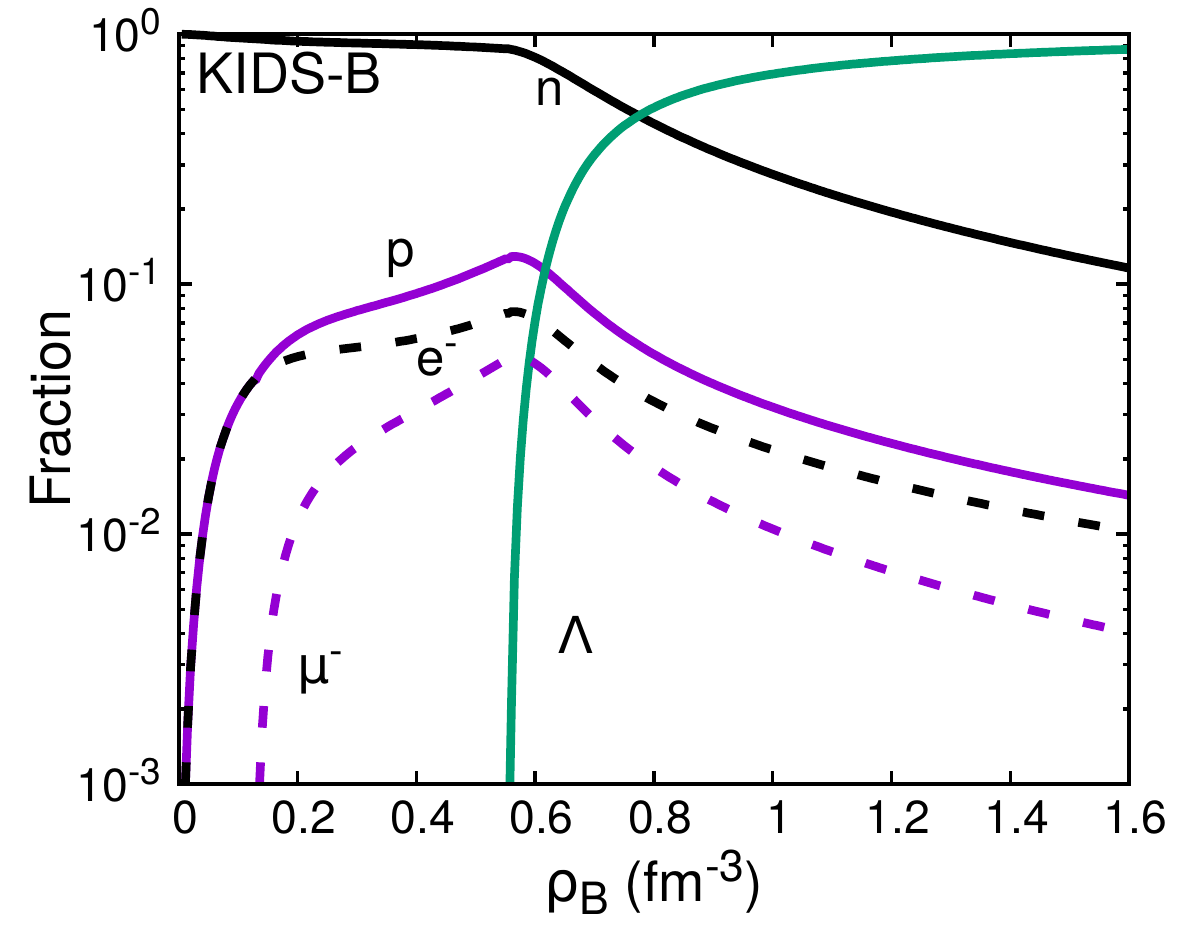}
\includegraphics[width=7cm]{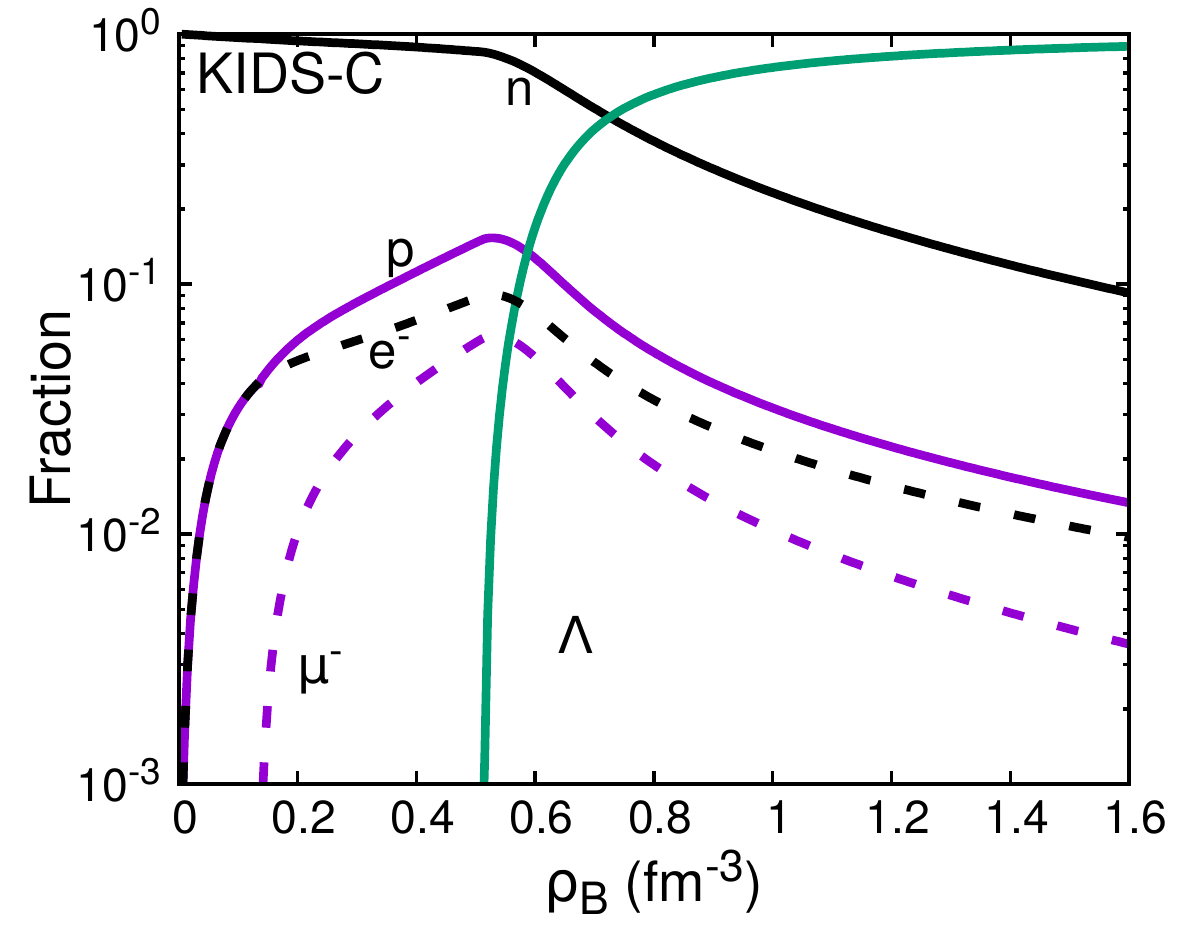}
\includegraphics[width=7cm]{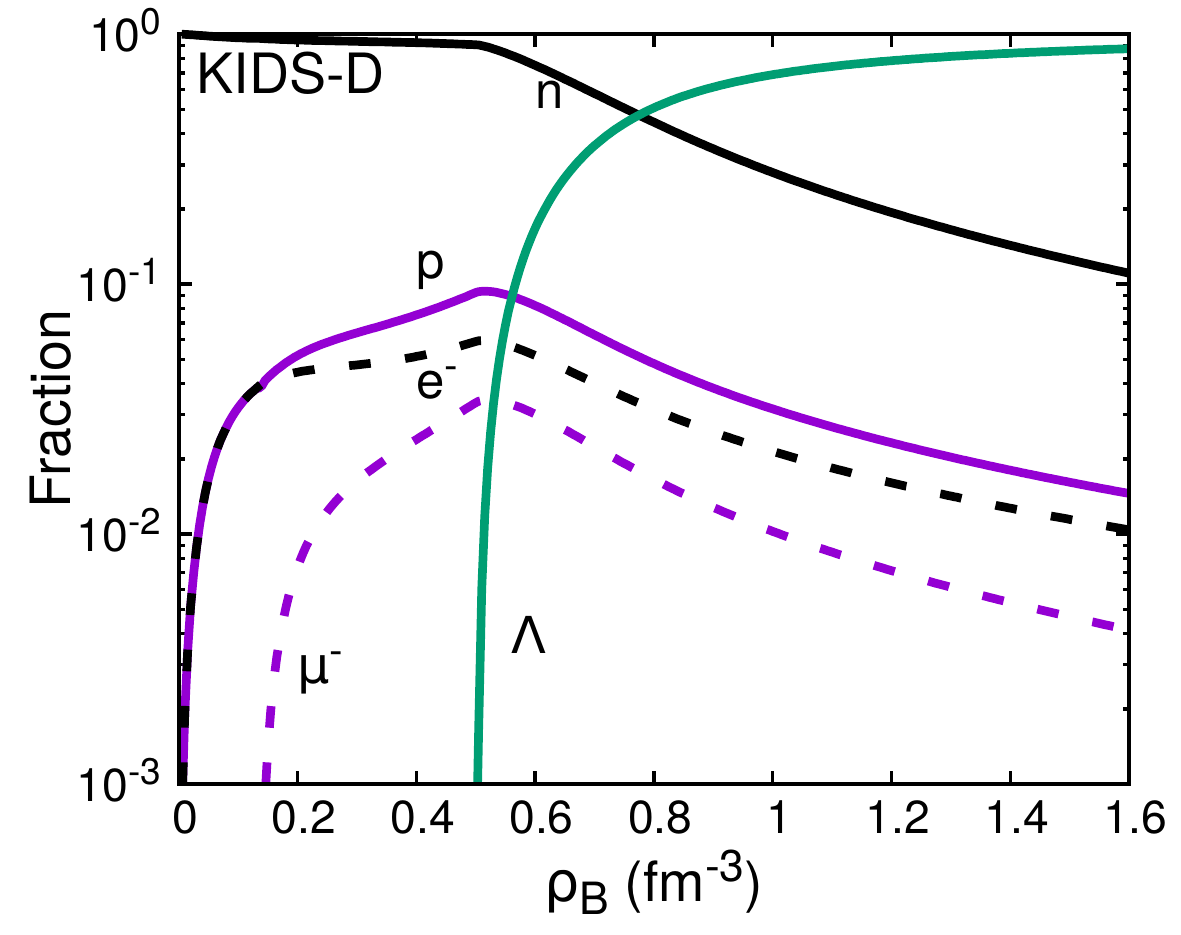}
\end{center}
\caption{Particle fractions in the core of neutron stars as a function of density
for the SLy4+HP$\Lambda2$, KIDS0, KIDS-A, KIDS-B, KIDS-C and KIDS-D models equipped with Y4 hyperon interaction.}
\label{fig7}
\end{figure}

Figure~\ref{fig7} shows the particle fraction defined as $\rho_i/\rho_{\rm B}$ where $\rho_i$ is the density of particle $i$ and $\rho_{\rm B} = \rho_n + \rho_p + \rho_\Lambda$ is the total baryon density.
SLy4+HP$\Lambda$2 model in Ref.~\cite{guleria2012} is included for comparison.
The density  $\rho_{\rm crit}$ at which $\Lambda$ starts to build up varies significantly.
In the unit of saturation density, $\rho_{\rm crit}$ are obtained as
2.8, 4.4, 3.5, 3.3 and 3.2 for SLy4, KIDS0, KIDS-B, KIDS-C and KIDS-D, respectively.
In the majority of models considering hyperons in the neutron star, $\rho_{\rm crit}$ is located in the range $(2-3) \rho_0$ \cite{ijmpe2015},
but the KIDS models consistently obtain the $\rho_{\rm crit}$ larger than $3 \rho_0$.
Even no hyperon creation is predicted in the KIDS-A model.
In Ref.~\cite{epja2020}, two-body $\Lambda N$ and three-body $\Lambda NN$ forces are calculated with chiral effective field theory (EFT),
and the result is applied to the creation of $\Lambda$ hyperons in the core of the neutron star.
Conclusion of Ref.~\cite{epja2020} was that $\Lambda$ is not created in the neutron star.
Looking into the behavior of chemical potentials,
chiral EFT obtains the result very similar to that of KIDS-A model.
Details will be discussed with the chemical potential. 

Production rate of $\Lambda$ also depends on the model.
SLy4 model shows relatively slow increase of the number of $\Lambda$ hyperon compared to the KIDS model.
Among the KIDS models, B, C, D models show similar behavior, but the population of $\Lambda$ increases
explosively right after it is created in the KIDS0 model.
The different results in the particle fraction such as $\rho_{\rm crit}$ and production ratio can be understood well in terms of the chemical potential.

\begin{figure}
\begin{center}
\includegraphics[width=7cm]{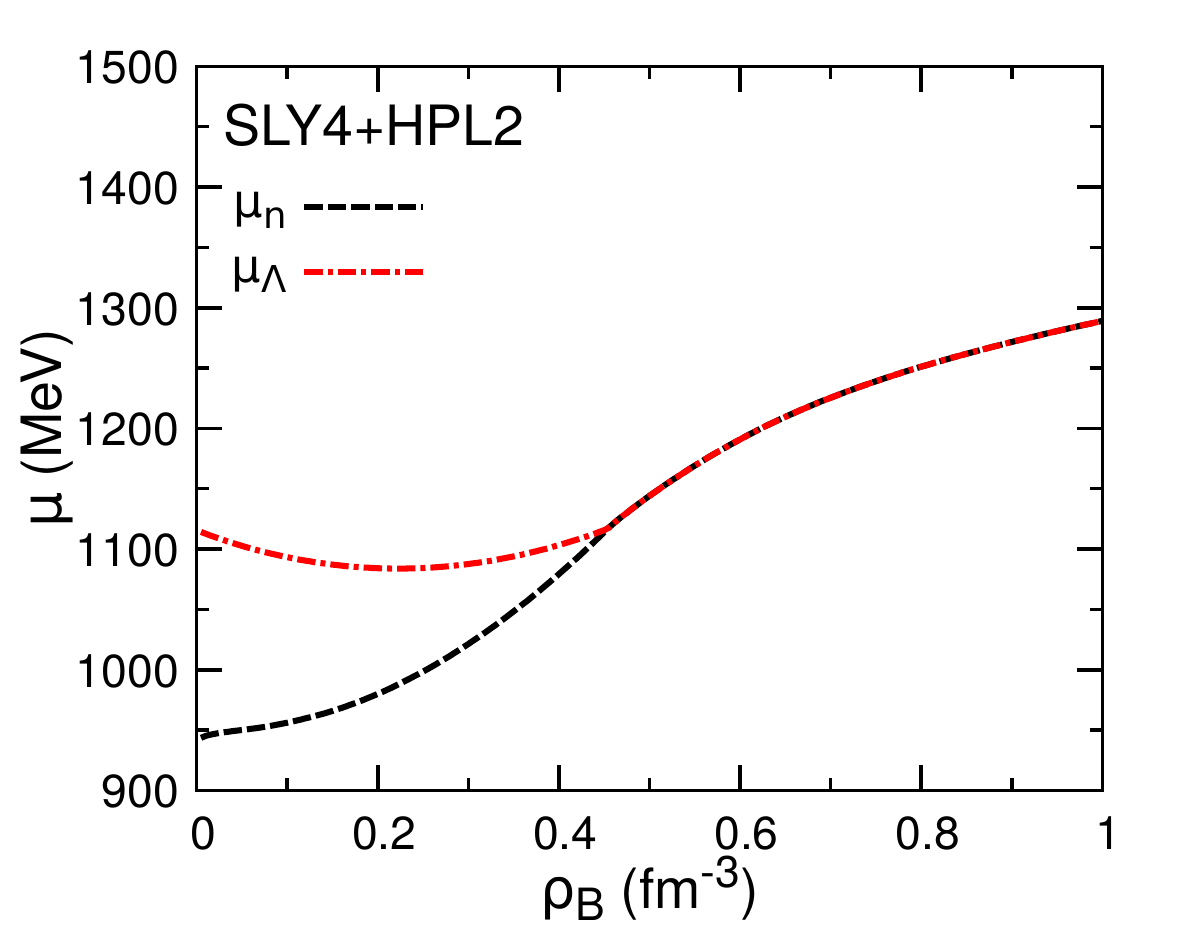}
\includegraphics[width=7cm]{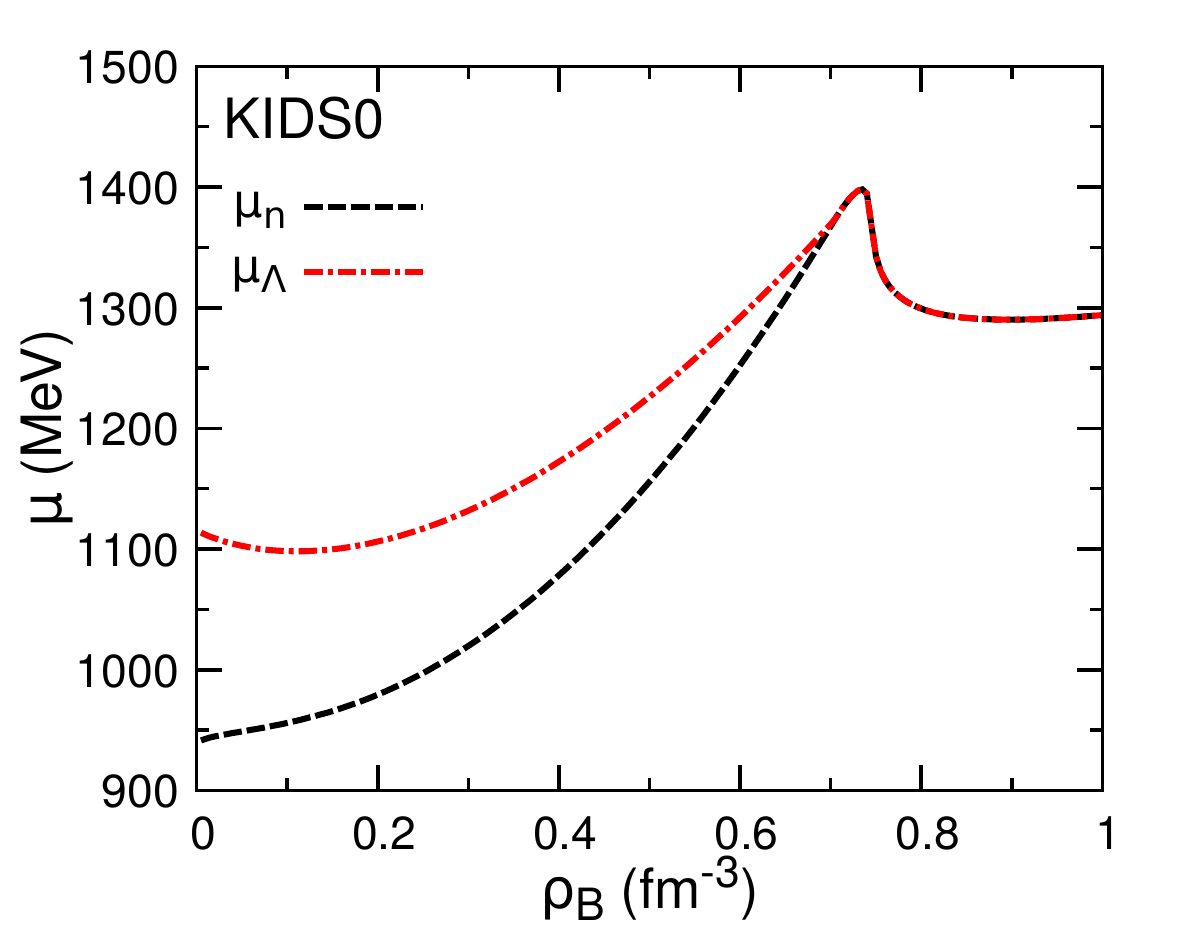}
\includegraphics[width=7cm]{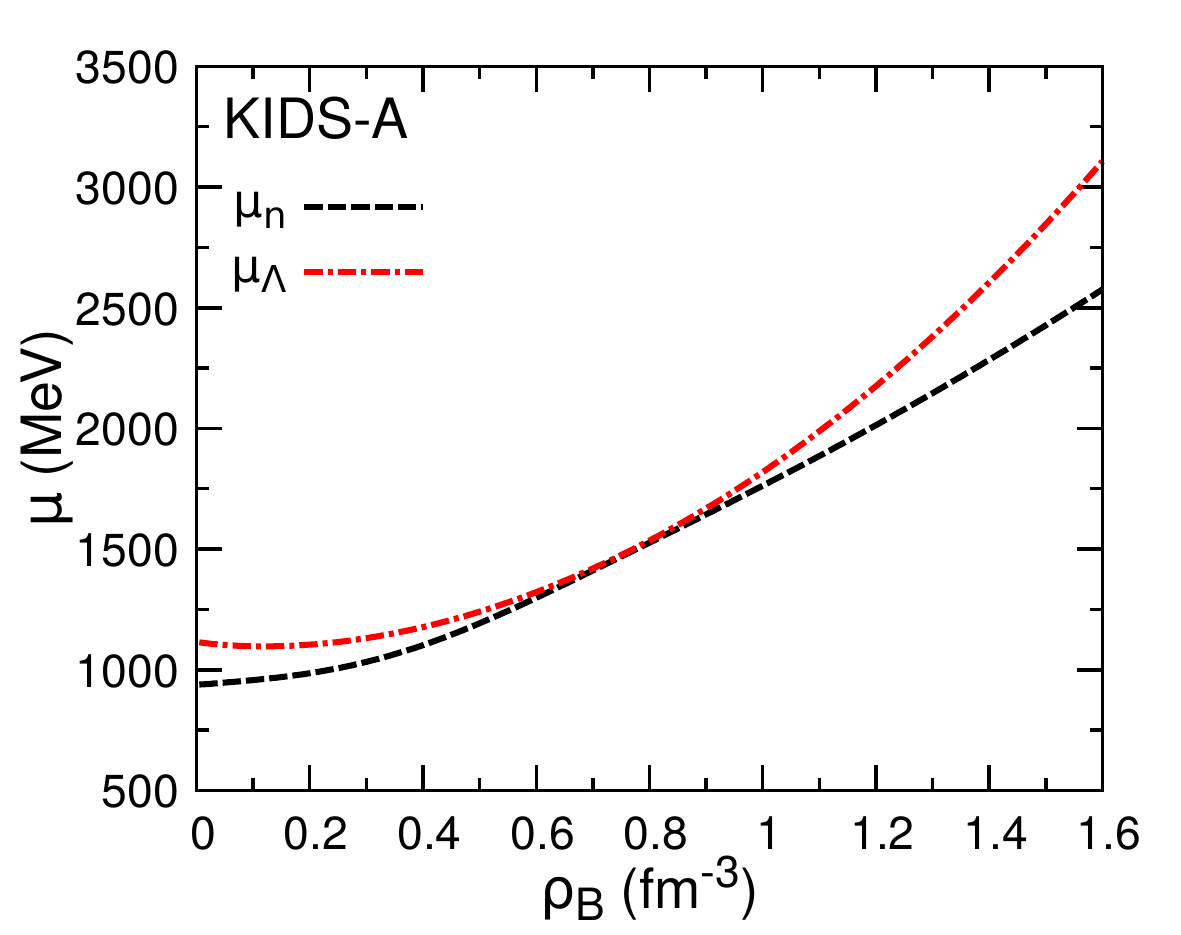}
\includegraphics[width=7cm]{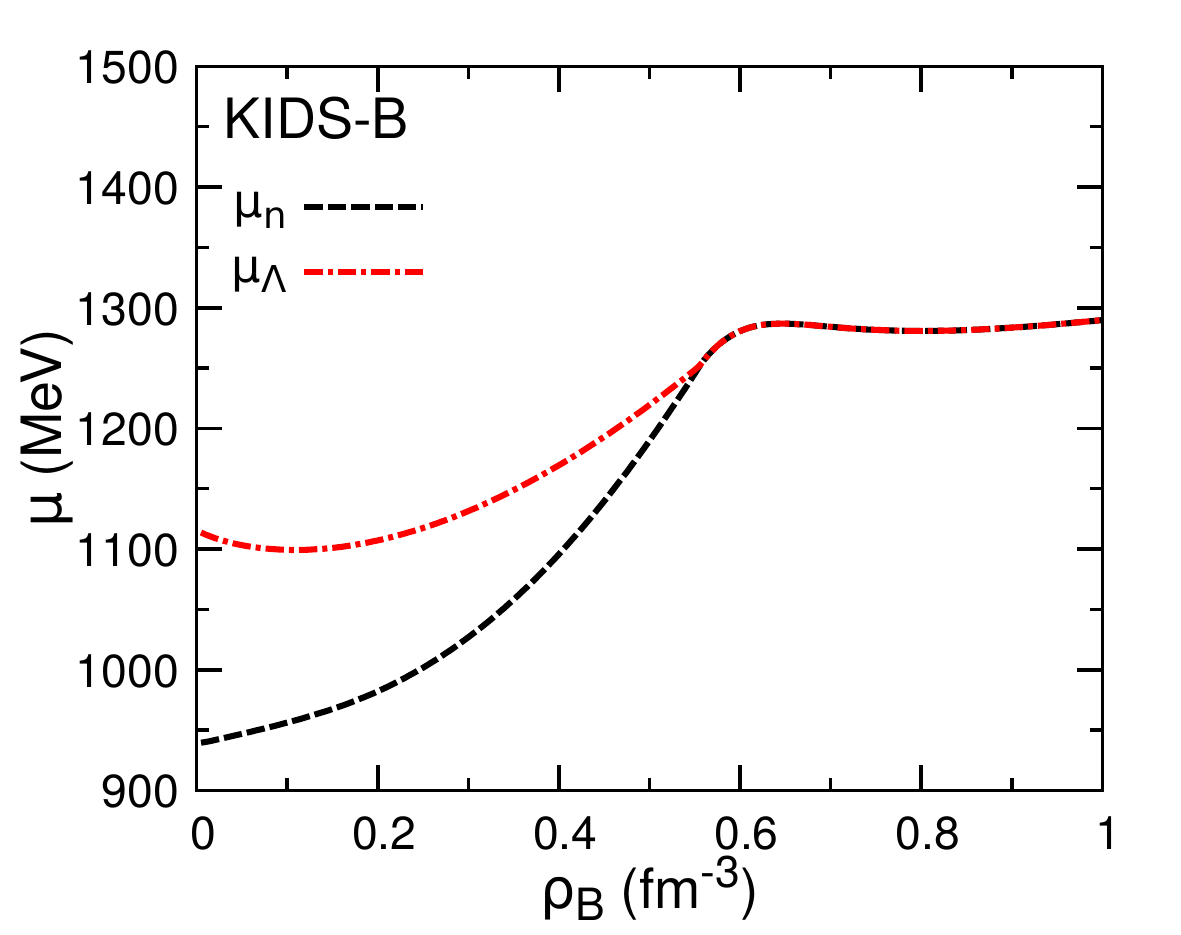}
\includegraphics[width=7cm]{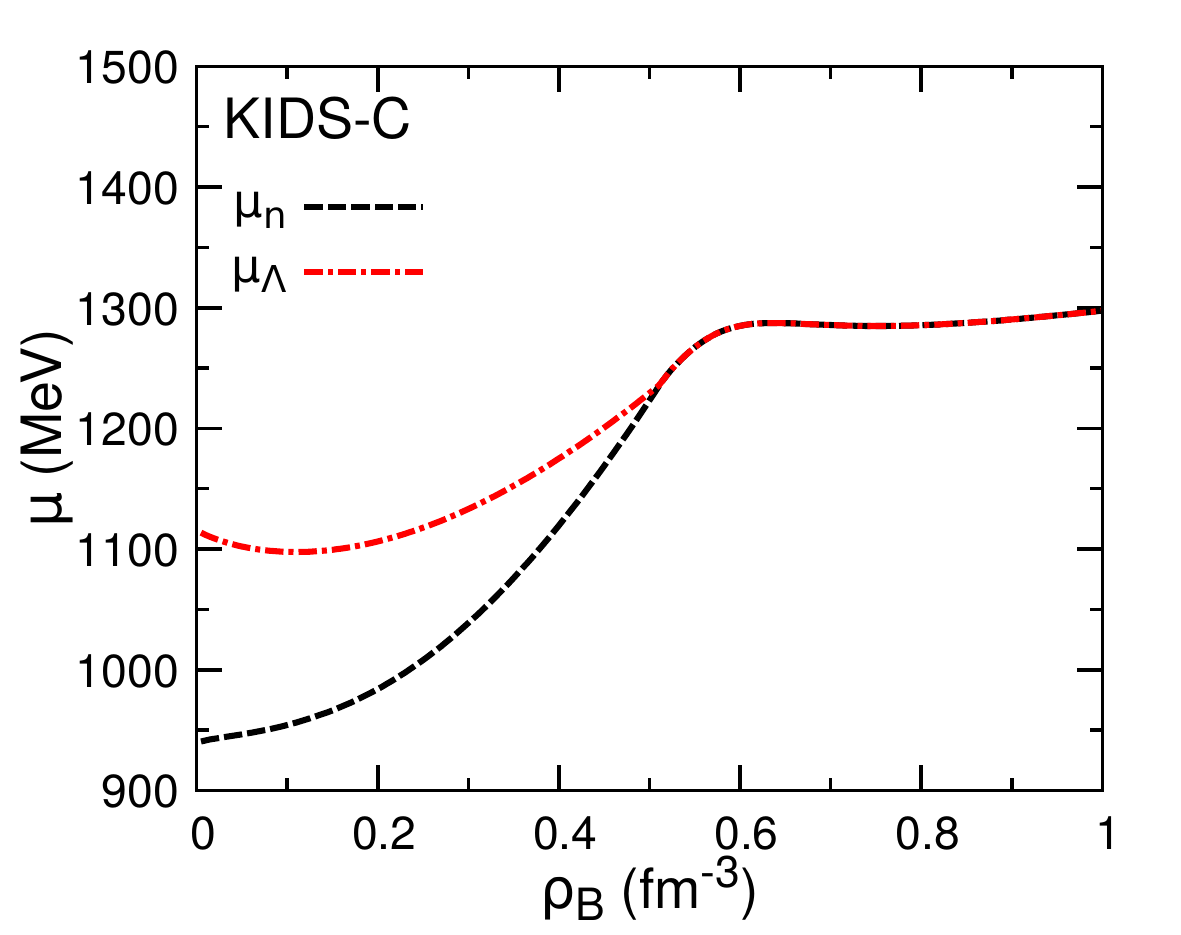}
\includegraphics[width=7cm]{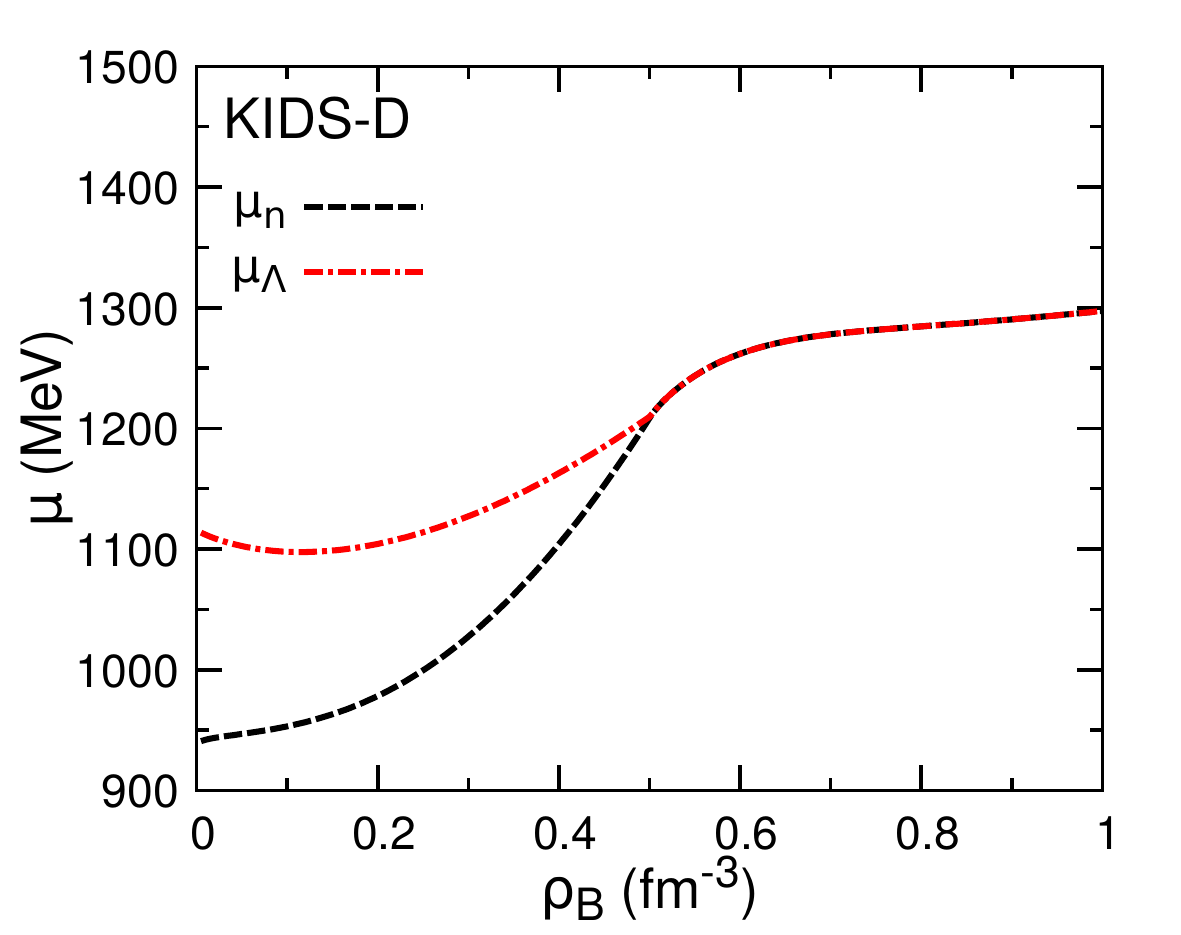}
\end{center}
\caption{Chemical potentials of the neutron and the $\Lambda$ hyperon for each model.}
\label{fig8}
\end{figure}

Figure \ref{fig8} shows the chemical potential of the neutron and $\Lambda$ hyperon in the core of a neutron star as functions of density.
All the models show that $\mu_n$ increases monotonically before the chemical equilibrium condition is satisfied.
$\mu_\Lambda$ is in decreasing phase up to 0.2 fm$^{-3}$ in the SLy4+HP$\Lambda$2 model, but the KIDS models equipped with Y4 interaction show decreasing behavior up to about 0.1 fm$^{-3}$, and change to increasing phase afterwards.
Hyperon creation density $\rho_{\rm crit}$ becomes small when $\mu_n$ increases fast 
or $\mu_\Lambda$ increases slowly.
Early creation of $\Lambda$ hyperons in the SLy4+HP$\Lambda$2 model is mainly originated from slow increase of $\mu_\Lambda$
relative to the KIDS models.
Large $\rho_{\rm crit}$ in the KIDS0 model, on the other hand, is because of the slow increase of $\mu_n$ compared to other KIDS models.
For example, at $\rho=0.6\, {\rm fm}^{-3}$, $\mu_n$ in the KIDS0 model is about 1250 MeV, but it is about 1300 MeV in the KIDS-A model.

Equation of state of nucleonic matter of the SLy4 model is very similar to that of the KIDS0 model \cite{Kim:2020tpj}, so the comparison of the two models can exhibit clearly the difference arising from hyperons.
In the KIDS-Y4 models, $\mu_\Lambda$ is increasing faster than the HP$\Lambda$2 model.
This difference in the behavior of $\mu_\Lambda$ makes huge difference in $\rho_{\rm crit}$ between SLy4+HP$\Lambda$2 and KIDS0-Y4.
Fast increase of $\mu_\Lambda$ in the Y4 models is mainly due to the contribution from the higher order terms
in the multiple density-dependent description of interactions.
Contribution of the multiple density dependence is essential in reproducing the hypernuclear data accurately and obtaining
stability in the fitting, but also critical in approaching to a complete description of neutron star EoS at supra densities.

The results of KIDS-A model, which are presented in scales of $x$ and $y$ axes larger than other models, show the non-existence of $\Lambda$ hyperon.
At densities $\rho \leq 2 \rho_0$, $\mu_n$ is substantially smaller than $\mu_\Lambda$.
As the density passes through $2\rho_0$, $\mu_n$ approaches to $\mu_\Lambda$, 
and the two chemical potentials become very close from $3\rho_0$ to $6\rho_0$.
Above $6\rho_0$, increase rate of $\mu_\Lambda$ overwhelms that of $\mu_n$,
so there is no chance for $\Lambda$ hyperon to inhabit in the neutron star.
The overall behavior of $\mu_n$ and $\mu_\Lambda$ such as substantial difference below $2\rho_0$, and similar values in $(3-5)\rho_0$
resembles the results of chiral EFT \cite{epja2020}.

$\mu_n$ of KIDS-B and D models behave similarly at $\rho \leq 5 \rho_0$,
and $\mu_n$ of KIDS-C model is slightly stiffer than B and D models.
On the other hand, $\mu_\Lambda$ of KIDS-B and C models increases at similar rate before $\Lambda$ hyperon is created,
and $\mu_\Lambda$ of KIDS-D model is softer than the B and C models.
$\rho_{\rm crit}$ is lowered when $\mu_n$ is stiff or $\mu_\Lambda$ is soft.
Since $\mu_n$(KIDS-C) $> \mu_n$(KIDS-B), and $\mu_\Lambda$(KIDS-D) $< \mu_\Lambda$(KIDS-B),
$\Lambda$ hyperons are created in the KIDS-C and D models at densities smaller than the KIDS-B model.
In a nutshell, KIDS-Y4 models illustrate the high threshold density for the $\Lambda$ appearance compared to other models including hyperons.

\section{Summary and Conclusion}

The work was stimulated by the motivation
i) Determine the in-medium interaction of $\Lambda$ hyperon accurately,
ii) Understand the effect of the uncertainty in the symmetry energy to the in-medium $\Lambda$-$N$ interaction.

In order to achieve the first goal, we employed the KIDS density functional formalism.
Many-body contributions to the $\Lambda$-$N$ interactions are expanded in the power of $\rho^{1/3}$,
and parameters are fitted to hypernuclear data.
Improvement of fitting is examined by varying the number of density-dependent terms in the many-body interaction.
Fitting becomes accurate 
with larger number of terms as expected, but the accuracy is increased stepwise.
With a single density dependence, fitting result is poor indeed, so a few input data could not be reproduced
within the experimental error.
Accuracy of fitting is improved greatly when two or more density-dependent terms are considered.
Mean deviation (MD) values are similar for the two- and three-term fittings, 
and it reduces by a factor of about 2 with the four terms.
Fitting with five terms gives only marginal improvement, 
so we concluded that fitting is sufficiently accurate with four density-dependent
$\Lambda$ interactions with many nucleons.
We confirmed that, with the four terms, all the data are reproduced within the experimental error bars, and five-term result is almost 
identical to that of the four-term functional.

The second topic has been explored by using KIDS0, KIDS-A, B, C, and D models, which are constrained by nuclear data and neutron star observations,
and have different density dependence of the symmetry energy \cite{kids-k0}.
Parameters in the $\Lambda$-$N$ interactions for each nuclear model are fitted with the number of density-dependent 
many-body interaction terms fixed up to fourth order.
We obtained similar values of MD for all the models.
The models are applied to the calculation of energy levels that are not included in the fitting.
Single particle energy levels of heavy nuclei agree well with the data at high orbital states such as $f$ and $g$ states,
but the levels in the $s$ state in light nuclei are consistently smaller than the experimental data.
Deviation from the light nuclei data is 10~\% or less in most cases.
Consistent description of both light and heavy $\Lambda$-hypernuclei will be investigated in a future work.
As a reference for the future experiment, we apply the five KIDS models to
the binding energy of the $\Lambda$ hyperon in the single-$\Lambda$ isotopes $^A_\Lambda$Sn for $A=124-136$. 
We obtain the ground state binding energies in the range $24-25$ MeV.
The result is stable with respect to the variation in the density dependence of the symmetry energy.

It is shown that the dependence on the symmetry energy becomes transparent at supra-saturation densities.
The effect is probed by calculating the creation of $\Lambda$ hyperons in the neutron star matter.
We find that the density at which $\Lambda$ hyperons start to be created and its population after they are created
depend strongly on the symmetry energy.
The issue of the hyperon puzzle will be investigated thoroughly by considering the $\Lambda$-$\Lambda$ interactions
and the creation of $\Sigma$ and $\Xi$ hyperons in a subsequent work.

All in all, multiple density dependence in the energy density functional of $\Lambda$-$N$ interaction is not an optional choice,
but a mandatory condition for an accurate and predictive description of the properties of hypernuclei.

\section*{Acknowledgments}
The work of SC is supported by the Institute for Basic Science (IBS-R031-D1).
The work of EH was supported by JSPS KAKENHI Grant Numbers JP18H05407 and JP20H00155.
The work of CHH was supported by the National Research Foundation of Korea (NRF) grant
funded by the Korea government (No. 2020R1F1A1052495).
The work of MKC was supported by the National Research Foundation of Korea (NRF) grant
funded by the Korea government (No.2021R1A6A1A03043957 and No. 2020R1A2C3006177).

\end{document}